\documentclass[useAMS,usenatbib]{mn2e}
\usepackage[english]{babel}

\usepackage{fancyhdr}
\usepackage{amsfonts}
\usepackage{amsmath}
\usepackage{amssymb}
\usepackage{multicol}
\usepackage{layout}
\usepackage{graphicx}
\usepackage{epsfig}
\usepackage{verbatim}
\usepackage[T1]{fontenc}
 \usepackage{aecompl}
\usepackage{color}
\usepackage{multirow}

\usepackage{times}
\usepackage{natbib}

\newif\ifAMStwofonts
\AMStwofontstrue

\graphicspath{{./fig/}}

\title[Evolution of spherical overdensities in holographic dark energy]
{Evolution of spherical overdensities in holographic dark energy models}

\author[Naderi et~al.]{Tayebe Naderi$^{1}$, Mohammad Malekjani$^{1}$
\thanks{malekjani@basu.ac.ir} and
 Francesco Pace$^{2,3}$\\
$^1$ Department of Physics, Bu-Ali Sina University, Hamedan 65178, Iran\\
$^2$ Jodrell Bank Centre for Astrophysics, School of Physics and
Astronomy, The University of Manchester,
Manchester, M13 9PL, U.K.\\
$^3$ Institute of Cosmology and Gravitation, University of
Portsmouth, Dennis Sciama Building, Portsmouth, PO1 3FX, U.K.}

\date{Accepted ?, Received ?; in original form \today}

\begin{document}
\maketitle

\label{firstpage}

\begin{abstract}
In this work we investigate the spherical collapse model in flat FRW
dark energy universes. We consider the Holographic Dark Energy (HDE)
model as a dynamical dark energy scenario with a slowly time-varying
equation-of-state (EoS) parameter $w_{\rm de}$ in order to evaluate
the effects of the dark energy component on structure formation in
the universe. We first calculate the evolution of density
perturbations in the linear regime for both phantom and quintessence
behavior of the HDE model and compare the results with standard
Einstein-de Sitter (EdS) and $\Lambda$CDM models. We then calculate
the evolution of two characterizing parameters in the spherical
collapse model, i.e., the linear density threshold $\delta_{\rm c}$
and the virial overdensity parameter $\Delta_{\rm vir}$. We show
that in HDE cosmologies the growth factor $g(a)$ and the linear
overdensity parameter $\delta_{\rm c}$ fall behind the values for a
$\Lambda$CDM universe while the virial overdensity $\Delta_{\rm
vir}$ is larger in HDE models than in the $\Lambda$CDM model. We
also show that the ratio between the radius of the spherical
perturbations at the virialization and turn-around time is smaller
in HDE cosmologies than that predicted in a $\Lambda$CDM universe.
Hence the growth of structures starts earlier in HDE models than in
$\Lambda$CDM cosmologies and more concentrated objects can form in
this case. It has been shown that the non-vanishing surface pressure
leads to smaller virial radius and larger virial overdensity
$\Delta_{\rm vir}$. We compare the predicted number of halos in HDE
cosmologies and find out that in general this value is smaller than
for $\Lambda$CDM models at higher redshifts and we compare different
mass function prescriptions. Finally, we compare the results of the
HDE models with observations.
\end{abstract}


\section{Introduction}
Cosmic structures such as galaxies and galaxy clusters develop from
the gravitational collapse
\citep{Gunn1972,Press1974,White1978,Peebles1993,Sheth1999,Peacock1999,Barkana2001,Peebles2003,Ciardi2005,
Bromm2011} of primeval small density perturbations originated during
the inflationary era \citep{Starobinsky1980,Guth1981,Linde1990}. To
study the non-linear evolution of cosmic structures a popular
analytical model, the spherical collapse model, was first introduced
by \cite{Gunn1972} and extended and improved by several following
works
\citep{Fillmore1984,Bertschinger1985,Hoffman1985,Ryden1987,Avila-Reese1998,Subramanian2000,Ascasibar2004,
Williams2004}. Recently the formalism of the spherical collapse
model was extended to include shear and rotation
\citep{Popolo2013a,Popolo2013b,Popolo2013c} and non-minimally
coupled models \citep{Pace2014}. In the spherical collapse model, at
early times primordial spherical overdense regions expand along the
Hubble flow at early times and since the relative overdensity of the
overdense region with respect to the background is small, the linear
theory is able to follow their evolution. At a certain point,
gravity starts dominating and opposes the expansion slowing it down
till the sphere reaches a maximum radius and completely detaches
from the background expansion. The following phase is represented by
the collapse of the sphere under its own self-gravity. In the
approximations introduced by the model, the collapse ends only when
the final radius becomes null. This is obviously not the case with
real structures, as virialization takes place thanks to non-linear
processes converting the kinetic energy of collapse into random
motions. The exact process of collapse due to gravitational
instability depends strongly on the dynamics of the background
Hubble flow.

In the last two decades, the astronomical data from SNe Ia \citep{Riess1998,Perlmutter1999,Riess2004,Riess2007}, CMB
\citep{Jaffe2001,Ho2008,Komatsu2011,Jarosik2011,Planck2013_XV,Planck2013_XVI,Planck2013_XIX}, Large Scale Structure
(LSS) and Baryon Acoustic Oscillations (BAO) \citep{Tegmark2004a,Eisenstein2005,Percival2010} and X-ray
\citep{Allen2004,Vikhlinin2009} experiments indicate that the universe is expanding at an accelerated rate. In the
framework of General Relativity (GR), an exotic component with positive density and negative pressure, the so-called
Dark Energy (DE), is responsible for this accelerated expansion. Results of the Planck experiment
\citep{Planck2013_XVI} show that DE occupies about $68\%$, dark matter about $27\%$ and usual baryons occupy about
$5\%$ of the total energy budget of the universe.\\
DE not only affects the expansion rate of the background Hubble flow and the distance-redshift relation, but also
the scenario of structure formation. The main goal of this work is to study the effect of DE on structure formation
within the spherical collapse model framework.

The first and simplest model for DE is Einstein's cosmological constant with constant Equation-of-State (EoS)
parameter $w_{\Lambda}=-1$. For the cosmological constant, structure growth, both in linear and non-linear regimes,
has been discussed by \cite{Lahav1991,Lilje1992,Lacey1993,Eke1996b,Viana1996,Kulinich2003,Debnath2006,Meyer2012}.
However the cosmological constant suffers from the fine-tuning and cosmic coincidence problems
\citep{Sahni2000,Weinberg1989,Carroll2001,Peebles2003,Padmanabhan2003,Copeland2006}.
Structure growth has been also investigated in quintessence models with constant EoS parameter different from $-1$.
In quintessence models, the principal difference is that the energy density decreases with time, whereas for the
cosmological constant it remains constant throughout the cosmic history. For constant EoS parameters in the range
$-1<w_{\Lambda}<-1/3$, \cite{Horellou2005} showed that structures form earlier and are more concentrated in
quintessence than in $\Lambda$CDM models. In addition, the evolution of structure growth and cluster abundance in
quintessence DE models \citep{Wang1998,Lokas2001a,Lokas2001b,Basilakos2003,Mota2004} and chameleon scalar field
\citep{Brax2010} have been investigated. It has also been shown that predictions of the spherical collapse model
strongly depend on the scalar field potential adopted in a minimally coupled scalar field scenario \citep{Mota2004}.
When DE clusters, \cite{Basilakos2009,Basilakos2010} showed that more concentrated structures can be formed with
respect to a homogeneous DE model. \cite{Bartelmann2006} and \cite{Pace2010} extended the spherical
collapse model in the presence of early DE models and showed that the growth of structures is slowed down with
respect to the $\Lambda$CDM model. Hence to reach the same amplitude of fluctuations today, the structures have to
grow earlier in this type of models.

On the other hand, in recent years, a wealth of dynamical DE models
with a time-varying EoS has been proposed
\citep{Copeland2006,Li2011,Bamba2012}. Observationally, the latest
astronomical data from SNe Ia, CMB and BAO experiments show that
dynamical DE models with time-varying EoS parameter are mildly
favored
\citep{Zhao2012,Alam2004,Gong2005a,Gong2005b,Huterer2005,Wang2005}.
The Holographic Dark Energy (HDE) model ( see section \ref{sect:HDE}
for a complete description of the model) is one of the most
interesting proposal in the category of dynamical DE scenarios. In
this work we study the evolution of spherical overdensities in HDE
models. The HDE model is considered as a dynamical DE model with
time-varying EoS parameter which can dominate the Hubble flow and
influence the growth of structures in the Universe. Here we consider
the non-interacting case of HDE model. In this case, DE is minimally
coupled to dark matter and the energy density of DE and dark matter
is conserved separately. Therefore in non-interacting HDE models, we
assume a uniform distribution of DE inside the perturbed region. In
this case, the energy density of DE remains the same both inside and
outside the overdense region. However, in the case of interacting
models, the energy density of dark matter and DE is not conserved
separately and their coupling is non-minimal. Hence the DE component
can cluster in a similar fashion as dark matter does.

The plan of the paper is the following. In section~\ref{sect:HDE} we present HDE cosmologies. In
section~\ref{sect:LPT} we discuss the linear evolution of perturbations in HDE cosmology and in
section~\ref{sect:scm} we study the non-linear spherical collapse model. In section~\ref{sect:results} numerical
results and comparison with observations have been presented. Finally we conclude in section~\ref{sect:conclusions}.

\section{Cosmology with Holographic dark energy}\label{sect:HDE}
The HDE model is constructed based on the holographic principle in
quantum gravity scenario \citep{tHooft1993,Susskind1995}. While
almost all dynamical DE models with time varying EoS are purely
phenomenological (there is no theoretically-motivated model behind
them), the advantage of the HDE model is that it originates from a
fundamental principle in quantum gravity, therefore possesses some
features of an underlying theory of DE. According to the holographic
principle, the number of degrees of freedom of a finite-size system
should be finite and bounded by the area of its boundary
\citep{Cohen1999}. In this case the total energy of a system with
size $L$ should not exceed the mass of a black hole with same size,
i.e., $L^3\rho_{\Lambda}\leq LM_{Pl}^2$, where $\rho_{\Lambda}$ is
the quantum zero-point energy density caused by UV cut-off $\Lambda$
and $M_{Pl}$ is the Planck mass ($M_{Pl}=1/8\pi G$). In a
cosmological context, when the whole Universe is taken into account,
the vacuum energy related to the holographic principle is viewed as
DE, the so-called HDE. The largest IR cut-off $L$ is chosen by
turning the previous inequality into an equality, hence the
following equation is taken for DE density in holographic models
\begin{equation}\label{eqn:h1}
\rho_{\rm de}=3c^2M_{Pl}^2L^{-2}\;,
\end{equation}
where $c$ is a positive numerical constant and the coefficient $3$ is just for convenience. An interesting feature of
HDE is that it has a close connection with the space-time \citep{Ng2001,Arzano2007}.

From an observational point of view, HDE models have been constrained by various astronomical observation
\citep{Alam2004,Huang2004,Zhang2005,Wu2008,Ma2009}. Using recent observational data, the value of the holographic
parameter $c$ in a flat Universe was constrained to $c=0.815^{+0.179}_{-0.139}$
\citep{Enqvist2004,Gong2004,Huang2004a,Huang2004b,Li2009}. The cosmic coincidence problem can be solved by inflation
in HDE model \citep{Li2004}.

It should be noted that the HDE model is defined by assuming an IR cut-off $L$ in Eq.~(\ref{eqn:h1}). The simplest
choice for IR cut-off is the Hubble length, $L=H^{-1}$. If we take $L$ as the Hubble scale $H^{-1}$, DE density will
be close to the observational data. However, in this case we get a wrong EoS for HDE models and the current
accelerated expansion of the Universe can not be recovered \citep{Horava2000,Cataldo2001,Thomas2002,Hsu2004}.
Another choice for the IR cut-off is the particle horizon, which however does not lead to the current accelerated
expansion \citep{Horava2000,Cataldo2001,Thomas2002,Hsu2004}. The third choice for the IR cut-off is the future event
horizon which was first assumed by \cite{Li2004} for HDE models. The event horizon is given by
\begin{equation}\label{eqn:h2}
 R_{\rm h}=a \int^{\infty}_t{\frac{dt}{a(t)}}=a\int_t^{\infty}{\frac{da}{Ha(t)}}\;,
\end{equation}
where $a$ is the scale factor and $t$ is cosmic time. In the context of the event horizon, the HDE model can generate
the late time acceleration consistently with observations \citep{Pavon2005,Zimdahl2007,Sheykhi2011}. The coincidence
and fine-tuning problems are also solved in this case \citep{Li2004}. In fact, a time varying DE model results in a
better fit compared with the standard cosmological constant based on analysis of cosmological data of type Ia
supernova \citep{Alam2004,Gong2005a,Gong2005b,Huterer2005,Wang2005}.

The dynamics of a flat FRW Universe containing pressure-less dark matter and a DE components is given by Friedmann
equation as follows
\begin{equation}\label{eqn:fh3}
 H^{2}=\frac{1}{3M_{Pl}^{2}}(\rho_{\rm m}+\rho_{\rm de})\;,
\end{equation}
where $\rho_{\rm m}$ and $\rho_{\rm de}$ are the energy density of the pressure-less matter and DE components,
respectively, and $H$ is the Hubble parameter. Here we use Eq.~(\ref{eqn:h1}) for the energy density of DE. For
non-interacting dark energy models, the energy density of DE and dark matter is given by the following continuity
equations
\begin{eqnarray}
&&\dot{\rho}_{\rm m}+3H\rho_{\rm m}=0, \label{eqn:contmt}\\
&&{}\dot{\rho}_{\rm de}+3H(1+w_{\rm de})\rho_{\rm de}=0\;,\label{eqn:contdt}
\end{eqnarray}
where the dot is the derivative with respect to cosmic time and $w_{\rm de}$ is the EoS parameter of DE.\\
Taking the time derivative of Friedmann Eq.~(\ref{eqn:fh3}) and using Eqs.~(\ref{eqn:contmt}, \ref{eqn:contdt}),
the relation $\dot{R}_{\rm h}=1+HR_{\rm h}$ and also the expression for the energy density of HDE models
$\rho_{\rm de}=3c^2M_{Pl}^2R_{\rm h}^{-2}$, the EoS parameter of the HDE model is
\begin{equation}\label{eqn:eos}
w_{\rm de}=-\frac{1}{3}-\frac{2\sqrt{\Omega_{\rm de}}}{3c}\;,
\end{equation}
where $\Omega_{\rm de}$ is the dimensionless density parameter of the DE component.

At late times, when DE dominates the energy budget of the Universe ($\Omega_{\rm de}\rightarrow 1$), we obtain
$w_{\rm de}<-1$ for $c<1$. In this case the EoS parameter of the HDE model is in the phantom regime
($w_{\rm de}<-1$). For $c\geq1$ we get $-1\leq w_{\rm de}<-1/3$, indicating the quintessence regime. The analysis of
the properties of DE from some recent observations favor models with $w_{\rm de}$ crossing $-1$ in the near past
\citep{Zhao2012,Alam2004,Gong2005a,Gong2005b,Huterer2005,Wang2005}.
The evolution of the DE density parameter $\Omega_{\rm de}$ in HDE models can be obtained by taking the time
derivative of $\Omega_{\rm de}=\rho_{\rm de}/\rho_{\rm c}=1/(HR_{\rm h})^2$ as follows:
\begin{equation}\label{eqn:evol}
\Omega_{\rm de}^{\prime}=\Omega_{\rm de}(1-\Omega_{\rm de})\left(1+\frac{2\sqrt{\Omega_{\rm de}}}{c}\right)\;,
\end{equation}
where the prime is the derivative with respect to $x=\ln{a}$. Since $a=1/(1+z)$, where $z$ is the cosmic redshift,
we have $d/dt=Hd/dx=-H(1+z)d/dz$. In terms of the cosmic redshift, Eq.~(\ref{eqn:evol}) is written as
\begin{equation}\label{eqn:evolz}
\frac{d\Omega_{\rm de}}{dz}=-\frac{1}{(1+z)}\Omega_{\rm de}(1-\Omega_{\rm de})
\left(1+\frac{2\sqrt{\Omega_{\rm de}}}{c}\right)\;.
\end{equation}
Also, the differential equation for the evolution of the dimensionless Hubble parameter, $E(z)=H/H_0$, in HDE model
can be obtained by taking a time derivative of the Friedmann Eq.~(\ref{eqn:fh3}) and using
relations~(\ref{eqn:contmt}, \ref{eqn:contdt} and \ref{eqn:eos}) as follows:
\begin{equation}\label{eqn:hub1}
\frac{dE}{dz}=-\frac{1}{(1+z)}E\left(\frac{1}{2}\Omega_{\rm de}+\frac{\Omega_{\rm de}^{3/2}}{c}-\frac{3}{2}\right)\;.
\end{equation}

The system of coupled Eqs.~(\ref{eqn:eos}), (\ref{eqn:evolz}) and (\ref{eqn:hub1}) can be solved numerically to
obtain the evolution of the EoS parameter, energy density and Hubble parameter in HDE models as a function of cosmic
redshift. To fix the cosmology, we assume the present values of matter density and DE density parameters as:
$\Omega_{\rm m,0}=0.27$ and $\Omega_{\rm de,0}=0.73$ in a spatially flat Universe. The present Hubble parameter is
$H_0=70$~km/s/Mpc.

\begin{figure}
 \centering
 \includegraphics[width=7cm]{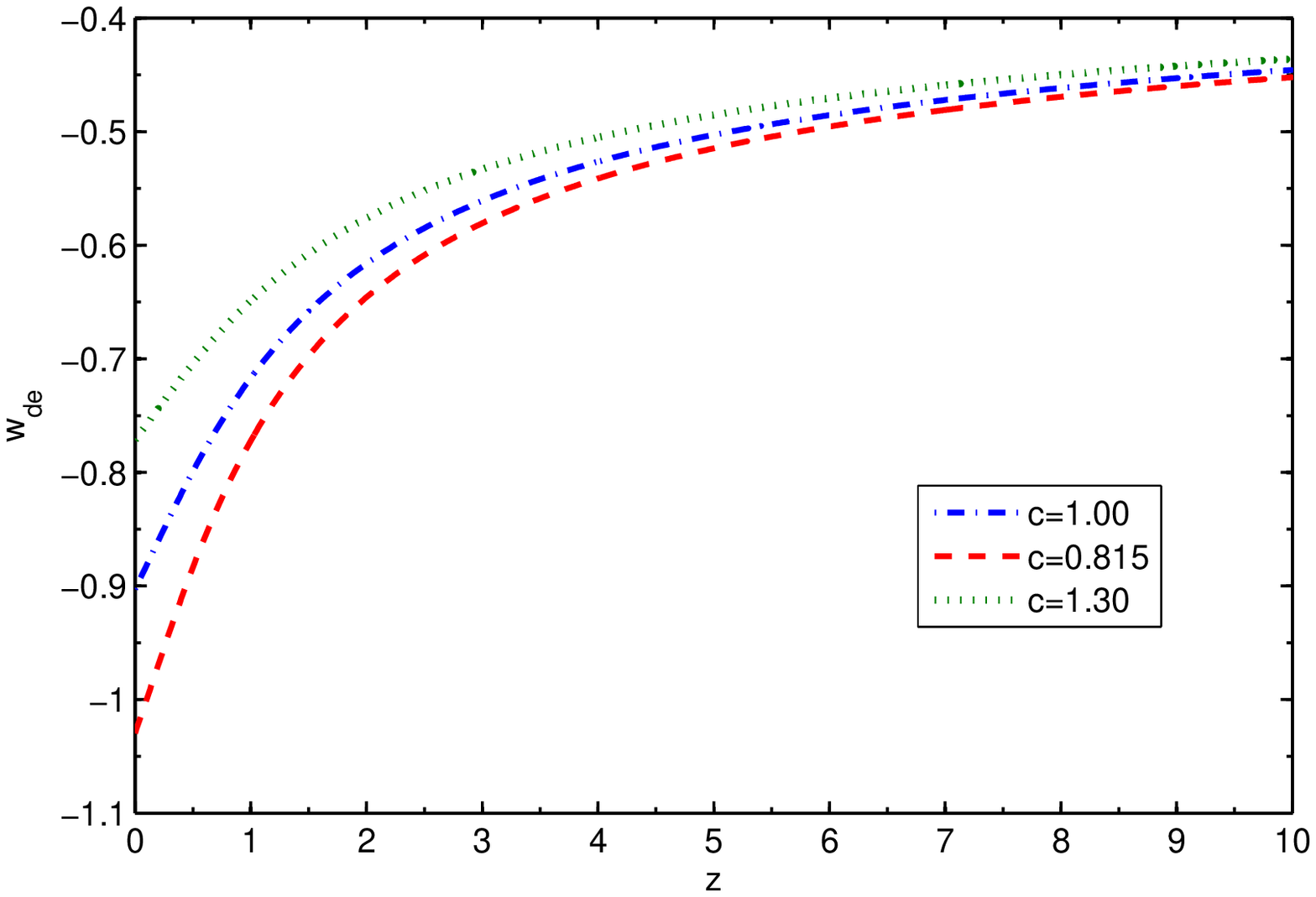}
 \includegraphics[width=7cm]{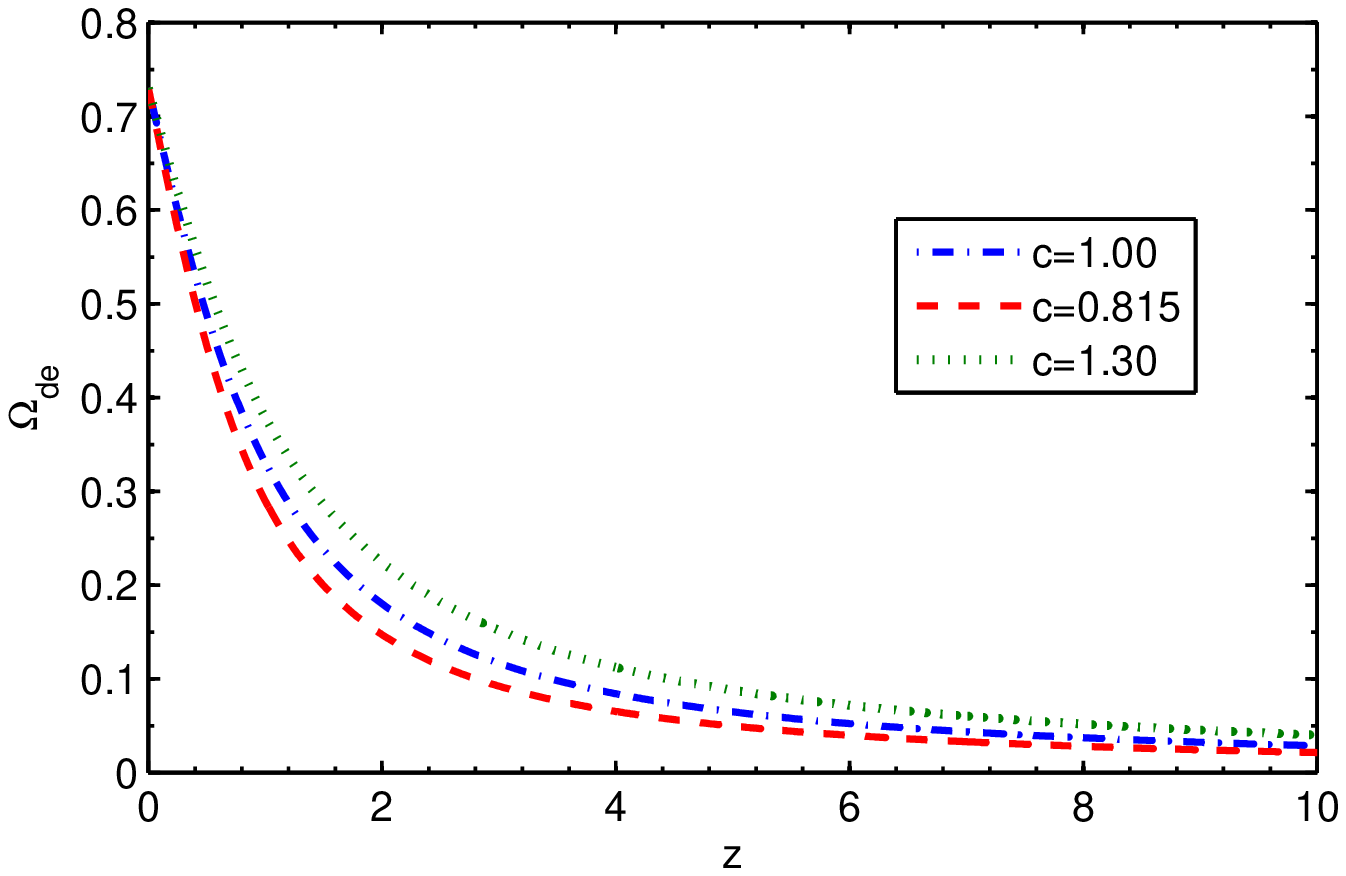}
 \includegraphics[width=7cm]{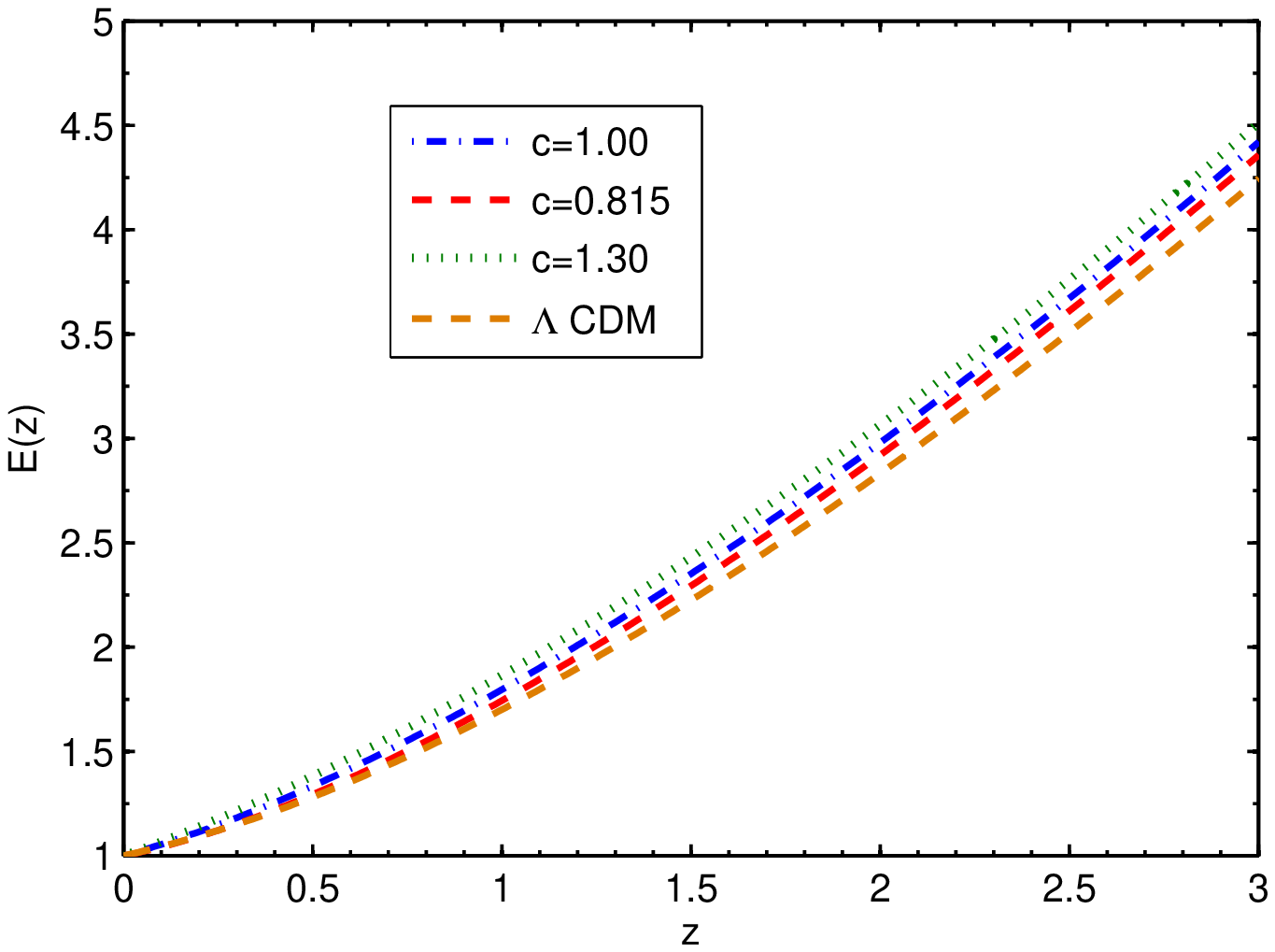}
 \caption{Top panel: Evolution of the EoS parameter of HDE model $w_{\rm de}$ as a function of cosmic redshift $z$
 for different values of the parameter $c$. Middle: Evolution of the DE density parameter $\Omega_{\rm de}$.
 Bottom: Evolution of the dimension-less Hubble parameter $E(z)=H/H_0$. The red dashed line represents the model with
 $c=0.815$, the blue dot-dashed curve the model with $c=1.00$ and the green-dotted line the model with $c=1.30$. For
 $c\geq 1$, the EoS parameter can not cross the phantom line $w=-1$ (quintessence regime). For the constrained value
 $c=0.815$ the phantom regime is achieved.}
 \label{fig:back}
\end{figure}

In figure (\ref{fig:back}) we show the evolution of the DE EoS parameter $w_{\rm de}$ (top panel), DE density
parameter $\Omega_{\rm de}$ (middle panel) and dimensionless Hubble parameter $E=H/H_0$ (bottom panel) as a function
of the cosmic redshift $z$ for three different values of the model parameter $c$. We see that for $c\geq 1$, the EoS
parameter can not enter into the phantom regime and remains in the quintessence regime. For $c<1$, we adopt the
constrained value $c=0.815$ from the observational data \citep{Alam2004}. In this case (red-dashed line) the phantom
regime can be crossed in the near past in agreement with observations
\citep{Alam2004,Gong2005a,Gong2005b,Huterer2005,Wang2005}.\\
The evolution of $\Omega_{\rm de}$ and $E(z)$ depends on the value of the parameter $c$. The Hubble parameter and DE
density are bigger in the quintessence regime (green-dotted line).

The cosmic time $t$ as a function of redshift $z$ is given by
\begin{equation}\label{eqn:cosmict}
 t=\frac{1}{H_0}\int_{z}^{\infty}\frac{dz}{(1+z)E(z)}\;,
\end{equation}
where $E(z)$ is the dimensionless Hubble parameter in HDE model. We use Eq.~(\ref{eqn:cosmict}) in order to study
the spherical collapse model in HDE cosmologies in Sect.~\ref{sect:scm}.

\section{Linear perturbation theory}\label{sect:LPT}
Here we study the linear growth of perturbations of non-relativistic dust matter by calculating the evolution of the
growth factor $g(a)$ in HDE cosmologies and compare it with the solution found for the EdS and $\Lambda$CDM models.

For a non-interacting HDE model, we assume that only pressure-less matter is perturbed and DE is uniformly
distributed. In this case the differential equation for the evolution of $g(a)$ is given by
\citep{Percival2005,Pace2010,Pace2012}
\begin{equation}\label{eqn:gfe}
g^{\prime\prime}(a)+\left(\frac{3}{a}+\frac{E^{\prime}(a)}{E(a)}\right)g^\prime(a)-
\frac{3}{2}\frac{\Omega_{m,0}}{a^5 E^2(a)}g(a)=0\;.
\end{equation}

To obtain the linear growth of structures in HDE cosmologies, we solve numerically Eq.~(\ref{eqn:gfe}) by using
Eq.~(\ref{eqn:hub1}) for the evolution of the Hubble parameter.\\
To evaluate the initial conditions, since we are in the linear regime, we assume that the linear growth factor has a
power law solution, $g(a)\propto a^n$, with $n$ to be evaluated at the initial time. We recall that for an EdS model,
$n=1$, while in general, in the presence of DE, $n\neq 1$. To evaluate the initial slope $n$, we insert the power law
solution into the differential equation describing the evolution of the linear growth factor and we solve the
second-order algebraic equation obtained.

\begin{figure}
 \centering
 \includegraphics[width=8cm]{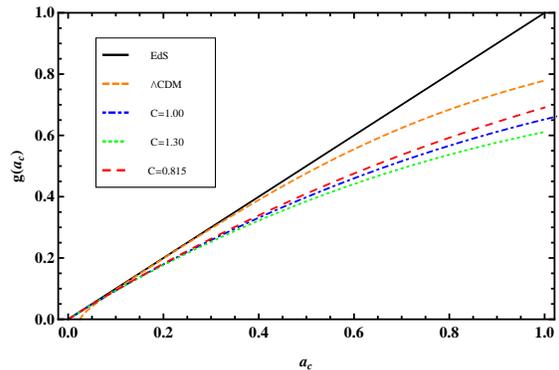}
 \caption{Time evolution of the growth factor as a function of the scale factor for the different cosmological
 models investigated in this work. Black solid line shows the EdS model, the orange short-dashed line the
 $\Lambda$CDM model, the blue (green) dot-dashed (dotted) line the HDE model with $c=1.00$ ($c=1.30$), while the
 red dashed line the HDE cosmology with $c=0.815$.}
 \label{fig:gf}
\end{figure}

In Fig.~(\ref{fig:gf}) we show the evolution of the linear growth factor $g(a)$ as a function of the scale factor.
We chose to normalize all the models to be the same at early times. We refer to the caption for line style and colors.
In the EdS model, the growth factor evolves proportionally to the scale factor, as expected. In the $\Lambda$CDM
model, the growth factor evolves more slowly compared to the EdS model since at late times the cosmological constant
dominates the energy budget of the universe. In the case of HDE model (phantom regime, $c<1$) with the constrained
holographic parameter $c=0.815$, $g(a)$ evolves more slowly than in the $\Lambda$CDM model. This is due to the fact
that the expansion of the Universe considerably slows down structure formation.

For the quintessence regime ($c\geq 1$), we notice that the evolution of $g(a)$ is smaller even when compared to
the phantom case. This behavior can be explained by taking into account the evolution of Hubble parameter in
Fig.~(\ref{fig:back}). The Hubble parameter is larger in the quintessence regime of the HDE models, it takes
intermediate values in the phantom regime and the smallest expansion appears to be in the $\Lambda$CDM model.
Therefore the growth factor $g(a)$ for the HDE will always fall behind the $\Lambda$CDM universe.

Here we conclude that different models for Hubble flow indicate different rate of structure growth in linear regime.
In the EdS Universe, in the absence of DE, the growth rate is largest. In the HDE model with $c=0.815$ the growth is
the smallest of the models here analyzed and it takes intermediate values for the $\Lambda$CDM universe. As a result
in the linear regime, we see that in HDE models the growth of structures is slowed down compared to $\Lambda$CDM and
EdS Universes due to bigger Hubble parameter. Hence to have the same fluctuations at the present time, perturbations
should start growing earlier in a HDE cosmology than in $\Lambda$CDM and EdS scenarios.

\section{Spherical collapse in HDE models}\label{sect:scm}
In this section we present the spherical collapse model in HDE cosmology. For this purpose, we first review the
basic equations we used to derive the correlation between turn-around and virial epochs and then obtain the virial
condition in this model. Finally we obtain the characteristic parameters of the spherical collapse model in the HDE
cosmologies.

In the scenario of structure formation, several attempts have been done to obtain the differential equation
governing the evolution of the matter perturbation $\delta$ in the limiting case of a matter dominated Universe
\citep{Bernardeau1994,Padmanabhan1996,Ohta2003,Ohta2004}. In the work of \cite{Abramo2007} the equation for the
evolution of $\delta$ was generalized to a universe containing a DE component with a time-dependent EoS. The
differential equation for the evolution of the overdensity $\delta$ in DE cosmologies and in the presence of
rotation and shear tensors has been derived in \cite{Popolo2013a,Popolo2013b}. For the case in which only the dark
matter component can cluster, the non-linear differential equation governing the evolution of $\delta$ is given by
\citep{Pace2010}
\begin{equation}\label{eqn:e1}
\delta^{\prime\prime}+\left(\frac{3}{a}+\frac{E^{\prime}(a)}{E(a)}\right)\delta^\prime-\frac{4}{3}
\frac{{\delta^\prime}^2}{1+\delta}-\frac{3}{2}\frac{\Omega_{m,0}}{a^5E^2(a)}\delta(1+\delta)=0\;,
\end{equation}
where the prime denotes the derivative with respect to the scale factor and $E$ is the dimensionless Hubble
parameter. In the linear regime, the above relation reduces to
\begin{equation}\label{eqn:e2}
\delta^{\prime\prime}+\left(\frac{3}{a}+\frac{E^{\prime}(a)}{E(a)}\right)\delta^\prime-
\frac{3}{2}\frac{\Omega_{m,0}}{a^5E^2(a)}\delta=0\;.
\end{equation}

\begin{figure}
 \centering
 \includegraphics[width=8cm]{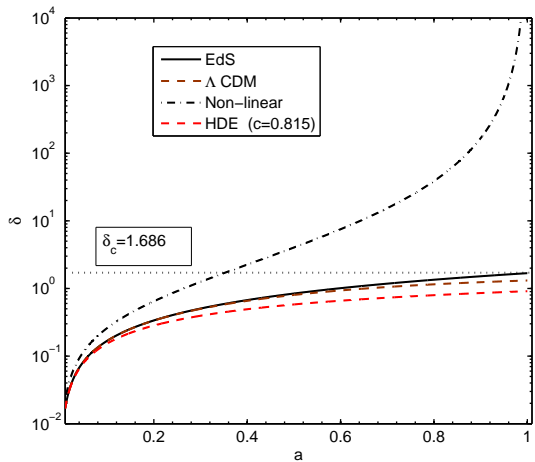}
 \caption{The growth of density perturbation $\delta$ in terms of the scale factor $a$ for different background
 models. The black dotted-dashed curve stands for non-linear evolution based on Eq.~(\ref{eqn:e1}). The black
 solid line indicates the linear growth of $\delta$ in CDM model. The brown dashed one is for the $\Lambda$CDM model
 and the red dashed curve represents the linear growth of $\delta$ in the phantom HDE background model.}
 \label{fig:delta}
\end{figure}

In Fig.~(\ref{fig:delta}), the growth of $\delta$ in the linear and
non-linear regimes is presented. To evaluate the initial conditions
for the differential equation describing the evolution of
perturbations, we refer to \cite{Pace2010} and following the
detailed procedure we set initial conditions to $\delta_{\rm
i}=2.09\times 10^{-4}$ and $\delta_{\rm i}^{\prime}=0$ at the
initial scale factor $a_{\rm i}=10^{-4}$. The black dotted-dashed
curve represents the solution for $\delta$ based on non-linear
evolution in Eq.~(\ref{eqn:e1}) for a collapse time $a=1$. The black
solid line is the linear evolution of $\delta$ for the EdS model
according to Eq.~(\ref{eqn:e2}). The brown and red dashed curves are
the linear evolution corresponding to the $\Lambda$CDM model and the
phantom regime of the HDE cosmology, respectively. We see that the
non-linear solution starts to deviate from the linear-based solution
and to grow very quickly. In the EdS model $\delta$ grows faster and
approaches the standard value $\delta_c=1.686$ at the collapse scale
factor. The linear growth for $\Lambda$CDM model is smaller compared
to the EdS scenario. In HDE model for $c=0.815$, $\delta$ grows more
slowly compared to the EdS and $\Lambda$CDM models.

\subsection{Turn-around and virial redshifts}
\begin{figure}
 \centering
 \includegraphics[width=8cm]{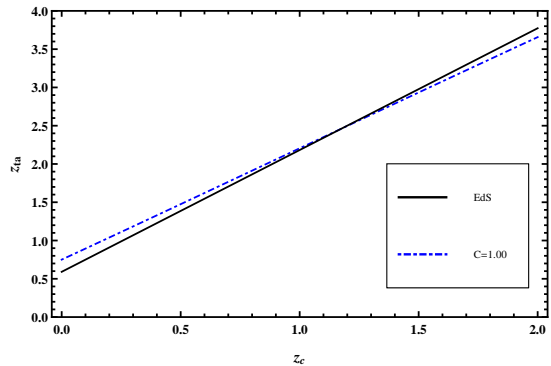}
 \caption{The turn-around redshift $z_{\rm ta}$ in terms of the collapse redshift $z_{\rm c}$ for HDE models.
 The solid curve stands for the EdS model and the dashed one corresponds to the HDE model with holographic parameter
 $c=1$.}
 \label{fig:zta}
\end{figure}

We know that in the EdS model the time for virialization in the spherical collapse model is twice the turn-around
time, i.e., $t_{\rm c}=2t_{\rm ta}$. Hence the ratio of the virial to the turn-around scale factor in the EdS
Universe is $a_{\rm c}/a_{\rm ta}=(t_{\rm c}/t_{\rm ta})^{2/3}=2^{2/3}=1.587$. Although this ratio is constant for
an EdS Universe, it changes in DE cosmologies. In HDE cosmologies, using Eq.~(\ref{eqn:cosmict}) for the evaluation
of the cosmic time we have
\begin{equation}
 \int_{z_{\rm c}}^{\infty}\frac{dz}{(1+z)E(z)}=2\int_{z_{\rm ta}}^{\infty}\frac{dz}{(1+z)E(z)}\;.
\end{equation}
By solving the above integrals, we can determine the correlation between turn-around redshift $z_{\rm ta}$ and
collapse redshift $z_{\rm c}$. In Fig.~(\ref{fig:zta}), this correlation is shown for a HDE model with holographic
parameter $c=1$ ( blue dashed line) and EdS model (solid line). Using a linear fitting method, we obtained the
following fitting formulas as a correlation between turn-around and collapse redshifts

\begin{eqnarray}\label{eqn:zt}
z_{\rm ta} & = & 1.542 z_{\rm c}+0.720\;, \quad (c=1) \nonumber \\
z_{\rm ta} & = & 1.551 z_{\rm c}+0.728\;, \quad (c=1.3) \nonumber \\
z_{\rm ta} & = & 1.538 z_{\rm c}+0.713\;, \quad (c=0.815) \nonumber \\
z_{\rm ta} & = & 1.535 z_{\rm c}+0.740\;, \quad (\Lambda{\rm CDM}) \nonumber \\
z_{\rm ta} & = & 2^{0.67}(1+z_{\rm c})-1\;, \quad ({\rm EdS})\;.
\end{eqnarray}

The result for HDE cosmologies is compatible with the fitting formula obtained in $\Lambda(t)$CDM cosmology, see
Eq.~34 of \citep{Basilakos2010}. As an example, consider a galaxy cluster virializing at the present time
$z_{\rm c}=0$. The turn-around epoch takes place at $z_{\rm ta}=0.720$ ($c=1$), $z_{\rm ta}=0.728$
($c=1.3$), $z_{\rm ta}=0.713$ ($c=0.815$), $z_{\rm ta}=0.740$ ($w_{\rm de}=-1$). It should be noted the in a EdS
model the turn-around redshift corresponding to $z_{\rm c}=0$ is $z_{\rm ta}=0.591$. One can conclude that the
turn-around epoch takes place earlier for $\Lambda$CDM cosmologies, intermediate times are typical of HDE models
(by increasing the holographic parameter $c$ the turn-around redshift increases accordingly) and later for the EdS
Universe. As a further example, consider the virialization process at the higher redshift $z_{\rm c}=1.6$ at which
the most distant cluster has been observed \citep{Papovich2010}. In this case the corresponding turn around redshift
is: $z_{\rm ta}=3.187$ ($c=1$), $z_{\rm ta}=3.209$ ($c=1.3$), $z_{\rm ta}=3.173$ ($c=0.815$), $z_{\rm ta}=3.196$
($w_{\rm de}=-1$) and $z_{\rm ta}=3.136$ for the EdS Universe. Here we see that the turn-around redshift calculated
in HDE and $\Lambda$CDM models tends to the critical value $z_{\rm ta}=3.136$ of the EdS Universe. Therefore at high
redshifts the influence of DE on the virialization process is negligible. This result is expected, because at large
redshifts the Universe is dominated by matter and all the models approach the EdS cosmology.

\subsection{Virial theorem}
Here we investigate the virial theorem in DE cosmologies. The virial
theorem relates the kinetic energy $T$ to a potential energy of the
form $U\propto R^n$ as $T=(n/2)U$, where the energies of the system
are averaged over time \citep{Landau1960,Lahav1991}. In the EdS
Universe the potential energy due to gravitational force is in the
form of $U_{\rm G}\propto R^{-1}$ and the virial condition is
$2T+U_{\rm G}=0$, where $T$ is the kinetic energy. In the case of
the cosmological constant cosmologies the virial theorem reads
$2T+U_G=2U_{\rm de}$, where $U_{\rm de}$ is the potential energy due
to the DE field \citep{Lahav1991}. The kinetic and potential
energies in spherical geometry are given by
\begin{eqnarray}
 T & = & \frac{1}{2}\int{u^2\rho_{\rm s}}dV\;,\\
 U_{\rm G} & = & -\frac{1}{2}G\iint{\frac{\rho_{\rm s}(r)\rho_{\rm s}(r^{\prime})}
 {\rvert r-r^{\prime}\rvert}dVdV^{\prime}}\;,\\
 U_{\rm de} & = & -\frac{1}{2}G\iint{\frac{\rho_{\rm s}(r)\rho_{\rm de}(r^{\prime})}
 {\rvert r-r^{\prime}\rvert}dVdV^{\prime}}\;,
\end{eqnarray}
where $u$ is the peculiar velocity of the fluid element inside the spherical region. For homogeneous DE,
$\rho_{\rm de}$ is uniform inside the collapsing sphere and its dynamics is the same of the background level.
Also for a top-hat profile, the distribution of matter inside the spherical collapse is uniform. In this case for
spherical mass fluctuations, the above potential energies become

\begin{eqnarray}\label{potentialg}
&& U_{\rm G} = -\frac{16\pi^2G}{3}\int_{0}^{R}r^4\rho_{\rm s}^2(r) dr=-\frac{3GM^2}{5R}\;,\\
\label{potentiald} \nonumber && U_{\rm de} =
-\frac{16\pi^2G}{3}\int_{0}^{R}r^4\rho_{\rm s}(r)\rho_{\rm de}(r)dr=
\nonumber\\
&&{} -\frac{4\pi G \rho_{\rm de,0}MR^2}{5}(1+3w_{\rm de})
e^{3\int_{0}^{z}\frac{1+w_{\rm de}(z)}{1+z}dz}\;,
\end{eqnarray}
where in Eq.~(\ref{potentiald}) we considered the homogeneous time
varying DE with EoS parameter $w_{\rm de}(z)$. Note that in the
definition of the potential energy for time varying HDE field in
Eq.~(\ref{potentiald}), the coefficient $(1+3w_{\rm de})$ leads to a
positive DE potential as well, therefore representing a repulsive
potential. Thanks to the sign of the coefficient $(1+3w_{\rm de})$,
we don't need to assume the coefficient $2$ in the right hand side
of virial theorem defined by \cite{Lahav1991}. Also, since the DE
potential is considered with positive sign, the virial theorem in
this case reads $2T+U_{\rm G}+U_{\rm de}=0$. For $w_{\rm de}=-1$,
Eq.~(\ref{potentiald}) reduces to $U_{\rm de}=\frac{\Lambda
M}{10}R^2$ as in \citep{Basilakos2010}.

\subsection{Spherical collapse parameters}
Consider a spherical overdense region with uniform matter density (top-hat profile) $\rho_{\rm s}$ and radius $R$
embedded in a Universe described by its background dynamics elsewhere except for the perturbed region. The dynamics
of background follows from Friedmann equation~(\ref{eqn:fh3}). For non-interacting HDE models, DE does not cluster
and its energy density $\rho_{\rm de}$ remains the same both inside and outside the over-dense patch. Based on
Birkhoffs theorem in GR, the gravitational field inside a spherical symmetric shell should vanish, and this is in
agreement with Newtonian gravity. This allows us to use Newtonian dynamics to study the evolution of matter density
perturbations on scales much smaller than the horizon. Hence the dynamics of the perturbed region is given by
\begin{equation}\label{eqn:over1}
\frac{\ddot{R}}{R}=-4\pi G\left(p_{\rm de}+\frac{\rho_{\rm de}+\rho_{\rm s}}{3}\right)=
-4 \pi G\left[(w_{\rm de}+\frac{1}{3})\rho_{\rm de}+\frac{1}{3}\rho_{\rm s}\right]\;,
\end{equation}
where the overdot indicates a derivative with respect to the cosmic time. Here we use the EoS of DE component to
obtain the second equality in Eq.~(\ref{eqn:over1}). We know that at early times, when the overdensity of this region
is small enough, the expansion of the patch follows the Hubble flow and density perturbations grow (approximately)
linearly with the scale factor. With the increase of the density perturbation, the expansion of the perturbed region
detaches from the Hubble flow and its expansion velocity decreases.
Finally at a characteristic scale factor $a_{\rm ta}$, it completely detaches from the general expansion and starts
to collapse under its own gravitational field till virialization takes place. We call $z_{\rm ta}$ and $z_{\rm c}$
the redshifts corresponding to the turn-around and virialization epochs, respectively, and $R_{\rm ta}$ and
$R_{\rm c}$ are the corresponding maximum and virial radii, respectively. By defining the dimensionless parameters
$x=a/a_{\rm ta}$ and $y=R/R_{\rm ta}$, the evolution of the scale factor of the background and the overdense spherical
region (i.e., equations \ref{eqn:fh3} and \ref{eqn:over1}) are governed by the following equations, respectively:
\begin{eqnarray}
 \left(\frac{\dot{x}}{x}\right)^2 & = & H_{\rm ta}^2\Omega_{\rm mt}
 \left(x^{-3}+\frac{\rho_{\rm de}}{\rho_{\rm mt}}\right)\;,\label{eqn:new1}\\
 \frac{\ddot{y}}{y}& = & -\frac{H_{\rm ta}^2\Omega_{\rm mt}}{2}\left(\frac{\xi}{y^3}+
 (1+3w_{\rm de})\frac{\rho_{\rm de}}{\rho_{\rm mt}}\right)\;,\label{eqn:new2}
\end{eqnarray}
where $H_{\rm ta}$, $\rho_{\rm mt}$ and $\Omega_{\rm mt}$ are the Hubble parameter, the matter density and the
matter density parameter at turn-around time, respectively, and $\xi$ is the ratio of matter density inside the
sphere to the matter density at background at turnaround epoch $\xi=(\rho_{\rm s}/\rho_{\rm m})_{x=1}$. In order to
obtain Eq.~(\ref{eqn:new2}) we used the fact that the matter inside the sphere evolves as
\begin{equation}
 \rho_{\rm s}=\rho_{\rm st}\left(\frac{R}{R_{\rm ta}}\right)^{-3}=\frac{\xi\rho_{\rm mt}}{y^3}\;.
\end{equation}

We use the following fitting expression obtained by \cite{Wang1998,Lokas2001a,Lokas2001b,Basilakos2003,Mota2004}
in the line of COBE measurements
\begin{equation}\label{eqn:xi}
\xi=\left.\left(\frac{3\pi}{4}\right)^2\Omega_{m}^{-0.79+0.26\Omega_{m}-0.06w_{\rm de}}\right|_{x=1}\;.
\end{equation}
Following \citep{Wang1998,Lokas2001a,Lokas2001b,Basilakos2003,Mota2004}, we notice that in Eq.~(\ref{eqn:xi}) $\xi$
is weakly model dependent and can be used for models with time varying $w_{\rm de}$. We apply Eq.~(\ref{eqn:eos}) to
obtain $\xi$ in HDE models. In the limiting case of the EdS model, $\xi=5.6$ independently of cosmic time. Using
the virial theorem for DE cosmology $2T+U_{\rm G}+U_{\rm de}$=0 and the conservation equation between virial and turn
around epochs, $T_{\rm c}+U_{\rm G,c}+U_{\rm de,c}=U_{\rm G,t}+U_{\rm de,t}$, we obtain the following equation for
the dimensionless parameter $\lambda=R_{\rm c}/R_{\rm ta}$

\begin{eqnarray}\label{eqn:lambda}
&&\left[\Omega_{\rm de,0}(1+3w_{\rm de}(z_{\rm c}))A(z_{\rm c})\right]\lambda^3-
2\left[\xi(1+z_{\rm ta})^3\Omega_{\rm m,0}+\right.\nonumber\\
&&{}\left.\Omega_{\rm de,0}(1+3w_{\rm de}(z_{\rm ta}))A(z_{\rm ta})\right]\lambda+\xi(1+z_{\rm ta})^3
\Omega_{\rm m,0}\nonumber\\
&&{} =0\;,
\end{eqnarray}

where we used Eqs.~(\ref{potentialg}) and (\ref{potentiald}) and define the parameter
$A(z)=e^{3\int_{0}^{z}\frac{1+w_{\rm de}(z)}{1+z}dz}$. In the limiting case of an EdS Universe one can easily see
that $\lambda=1/2$, as expected.

In this framework the overdensity of the collapsing structure at the virialization time is given by
\begin{equation}
 \Delta_{\rm vir}=\frac{\rho_{\rm s,c}}{\rho_{\rm m,c}}=
 \frac{\xi}{\lambda^3}\left(\frac{1+z_{\rm ta}}{1+z_{\rm c}}\right)^3\;,
\end{equation}
where $(1+z_{\rm ta})/(1+z_{\rm c})$, $\xi$ and $\lambda$ are calculated from Equations~(\ref{eqn:zt}),
(\ref{eqn:xi}) and (\ref{eqn:lambda}), respectively. For completeness we also discuss the parameter $\delta_{\rm c}$
defined as the linear evolved primordial perturbations to the collapse epoch. The parameters $\delta_{\rm c}$ and
$\Delta_{\rm vir}$ are the two characterizing parameters in the spherical collapse model. In next section we present
our numerical results for HDE cosmologies and compare them with the concordance $\Lambda$CDM model as well as
observations.

The framework of the spherical collapse model can be extended to
include contributions from the shear and the angular momentum terms
in Eq.~(\ref{eqn:over1}, in complete analogy with
\cite{Popolo2013b}. While an extensive and quantitative analysis of
the effects of these two additional non-linear terms goes beyond the
purpose of this work, from recent works and based on physical
arguments, we can assert that we would expect similar results to
previous works, where, due to the additional mass dependence of the
spherical collapse parameters, low mass objects showed higher values
for the spherical collapse model parameters $\delta_{\rm c}$ and
$\Delta_{\rm vir}$, while high mass objects will be almost
completely unaffected. In particular we expect differences of the
order of several percent, based on results by \cite{Popolo2013b}.

 \subsection{Non-zero surface pressure} In this section we present
a short discussion on the inclusion of the non-vanishing surface
pressure term. The surface pressure is due to non-zero density at
the outer boundary of virialised clusters. A modified virial
relation in the presence of the surface pressure term is given by
\begin{equation}
2T+U_{\rm G}+U_{\rm de}=3P_{\rm ext}V,
\end{equation}
where $P_{\rm ext}$ is the pressure at the surface of virialised cluster and $V$ is the volume
\citep{Popolo2002,Afshordi2002}. The surface term is related to the total potential energy $U=U_{\rm G}+U_{\rm de}$
as $3P_{ext}V=-\nu U$ \citep{Afshordi2002}. Here we discuss how the non vanishing surface pressure can change the
spherical collapse parameters. For simplicity, we first assume the standard EdS cosmology. In this case, the modified
virial theorem reads $2T+(1+\nu)U_{\rm G}=0$ and the energy conservation between turn-around and virial epochs
results in $\lambda=(1-\nu)/2$. Hence for $\nu>0$, the parameter $\lambda$ is smaller than the standard value $1/2$
and $\Delta_{\rm vir}$ will be larger than standard EdS value $\sim 178$. For example, for a value $\nu=0.005$, one
obtains $\lambda=0.0.4975$ and $\Delta_{\rm vir}\simeq 182$ which is $2\%$ larger than the standard EdS value. In the
general case of a HDE universe, the modified virial theorem can be obtained as
$2T+(1+\nu)(U_{\rm G}+U_{\rm \Lambda})=0$. The condition for energy conservation between turn-around and virial
epochs, $T_{\rm c}+U_{\rm G,c}+U_{\rm de,c}=U_{\rm G,t}+U_{\rm de,t}$, as well as the modified virial condition lead
to the following cubic equation for $\lambda$:
\begin{eqnarray}\label{eqn:lambda2}
&&\Big[(1-\nu)\Omega_{\rm de,0}(1+3w_{\rm de}(z_{\rm c}))A(z_{\rm c})\Big]\lambda^3-2\Big[\xi(1+z_{\rm ta})^3
\nonumber\\
&&{} \Omega_{\rm m,0}+\Omega_{\rm de,0}(1+3w_{\rm de}(z_{\rm ta}))A(z_{\rm ta})\Big]\lambda+ \nonumber\\
&&{} (1-\nu)\xi(1+z_{\rm ta})^3\Omega_{\rm m,0} =0\;,
\end{eqnarray}
Inserting the value $\nu=0$ we recover Eq.~(\ref{eqn:lambda}) as
expected. In the next section we evaluate the effect of the
inclusion of the non-vanishing surface pressure on the spherical
collapse parameters $\lambda$ and $\Delta_{\rm vir}$ for the
illustrative value $\nu=0.005$ and holographic parameter $c=1$.

\subsection{Mass Function and Number density}
 We now calculate the comoving number density of virialised objects
in a given mass range. In the Press-Schechter formalism, the average
comoving number density of halos of mass M is described by the
universal mass function, $n(M, z)$:
\begin{equation}\label{eq:number_density}
 n(M,z)=\left(\frac{\bar{\rho}}{M^2}\right)\frac{d\log\nu}{d\log{M}} \nu f(\nu)\;,
\end{equation}
where $\bar{\rho}$ is the background density and $f(\nu)$ is the
so-called multiplicity function \citep{Press1974,Bond1991}. The
variable $\nu$ is defined as $\nu=\delta_c^2/\sigma^2(M)$, where
$\sigma(M)$ is the r.m.s. of the mass fluctuation in spheres of mass
$M$. Since both $\delta_c$ and $\sigma(M)$ evolve in time, also the
variable $\nu$ would be in principle a time-dependant function.
However, the linear overdensity parameter represents the initial
perturbation evolved via the linear growth factor, therefore $\nu$
is in reality time-independent and probes and initial amplitude of
perturbations. In a Gaussian density field, $\sigma$ is given by
\begin{equation}\label{eq:sigma}
\sigma^2(R)=\frac{1}{2\pi^2}\int_0^{\infty}{k^2P(k)}W^2(kR)dk\;,
\end{equation}
where $R=(3M/4\pi\rho_{m0})^{1/3}$ is the radius of the sphere at the present time,
$W(kR)=3[\sin(kR)-kR\cos(kR)]/(kR)^3$ is the Fourier transform of a spherical top-hat profile with radius $R$
and $P(k)$ is the power spectrum of density fluctuations \citep{Peebles1993}. The quantity $\sigma(M,z)$ can be
related to its present value as $\sigma(M,z)=g(z)\sigma(M,z=0)$, where $g(z)=\delta_c(z)/\delta_c(z=0)$ is the
linear growth factor. In this work, like \cite{Abramo2007}, we use the fitting formula given by \cite{Viana1996}
\begin{equation}\label{eq:sigma_variance}
\sigma(M,z)=\sigma_8(z)\left(\frac{M}{M_8}\right)^{-\gamma(M)/3}\;,
\end{equation}
where $M_8=6\times 10^{14}\Omega_{m0}h^{-1}M_{\odot}$ is the mass inside a sphere of radius $R_8=8h^{-1}$Mpc,
$\sigma_8$ is the mass variance of the overdensity on the scale of $R_8$ and the exponent $\gamma(M)$ depends on the
shape parameter $\Gamma=\Omega_{m0}h\exp(-\Omega_b-\Omega_b/\Omega_{m0})$. For a spectral index $n=1$ we have
\begin{equation}\label{eqn:gamma}
\gamma(M)=(0.3\Gamma+0.2)\left(2.92+\frac{1}{3}\log{\frac{M}{M_8}}\right)\;.
\end{equation}
 Being an approximation, Eqs.~\ref{eq:sigma_variance}
and~\ref{eqn:gamma} have a limited range of validity and as stated
in \cite{Viana1996} they match the true shape of $\sigma_8$ in a
region around $M_8$. For masses $M<M_8$ ($M>M_8$), the fitting
formula predicts higher (lower) values of the variance, with
differences at most of 10\%. This means that the mass function will
predict less (more) objects at lower (higher) masses. This is though
not an issue for our analysis, since all the models will be affected
in the same way and we are interested in the relative halo
abundance.

In order to determine the  predicted number density of dark matter
halos for a given cosmological model based on
Eq.~(\ref{eq:number_density}), we need the mass variance
$\sigma(M,z)$ and multiplicity function $f(\nu)$. The standard
function $f(\nu)=\sqrt{2/\pi} e^{-\nu/2}$ provides a good
representation of the observed distribution of virialised
structures. However, the standard mass function deviates from
simulations for low- and high-mass objects. Here, we use another
popular fitting formula proposed by Sheth \& Tormen
\citep{Sheth1999,Sheth2002}, the so-called ST mass function:
\begin{equation}\label{eq:multiplicity_ST}
 \nu f_{\rm ST}(\nu)= A_1 \sqrt{\frac{b\nu}{2\pi}}\left(1+\frac{1}{(b\nu)^p}\right)
\exp{\left(-\frac{b\nu}{2}\right)} \;,
\end{equation}
where the numerical parameters are: $A_1=0.3222$, $b=0.707$ and
$p=0.3$. For comparison, we also use the following improved mass
functions which are, respectively, presented in
\cite{Popolo2006a,Popolo2006b} (P06 mass function) and
\cite{Yahagi2004} (YNY mass function), respectively:
\begin{eqnarray}\label{eqn:multiplicity_popolo2006}
&&\nu f(\nu)=A_2\Big[1+\frac{0.1218}{(b\nu)^{0.585}}+
\frac{0.0079}{(b\nu)^{0.4}}\Big]\sqrt{\frac{b\nu}{2\pi}}\times\nonumber\\
&&{} \exp{\left(-0.4019 b\nu\left[1+\frac{0.5526}{(b\nu)^{0.585}}+
\frac{0.02}{(b\nu)^{0.4}}\right]^2\right)}\;,
\end{eqnarray}
\begin{eqnarray}\label{eqn:multiplicity_yahagi2004}
&&\nu f(\nu)=A_3\left[1+(B\sqrt{\nu/2})^C\right]\nu^{(D/2)}\times\nonumber\\
&&{} \exp{\{-(B\sqrt{\nu/2})^2\}}\;,
\end{eqnarray}
where $A_2=1.75$, $B=0.893$, $C=1.39$, and $D=0.408$, and
$A_3=0.298$.

To determine the  mass variance $\sigma$ via Eq.~(\ref{eq:sigma_variance}), we should first calculate $\sigma_8$.
The present value of $\sigma_8$ for HDE cosmologies is obtained with
\begin{equation}\label{eq:sigma_8}
\sigma_{8,\rm HDE}=\frac{\delta_{\rm c,HDE}(z=0)}{\delta_{c,\Lambda}(z=0)}\sigma_{8,\Lambda}\;,
\end{equation}
where $\sigma_{8,\Lambda}=0.8$ is the present value of $\sigma_8$ for the $\Lambda$CDM model. The value of
$\sigma_{8,\rm HDE}$ at redshift $z$ is therefore obtained as
$\sigma_{8,\rm HDE}(z)=g(z)\sigma_{8,\rm HDE}(z=0)$. Hence, the mass variance $\sigma(M,z)$ which is required to
compute the number density of halos can be easily obtained from Eq.~(\ref{eq:sigma_variance}). In the next section
we present the numerical results of the mass function formalism and number density clusters in HDE cosmologies.
We chose this normalization for the DE models studied here so to have approximately the same number of objects at
$z=0$. In this way differences will appear mainly at high redshifts.

\section{Numerical results and comparison with observation}\label{sect:results}
\begin{figure}
 \centering
 \includegraphics[width=8cm]{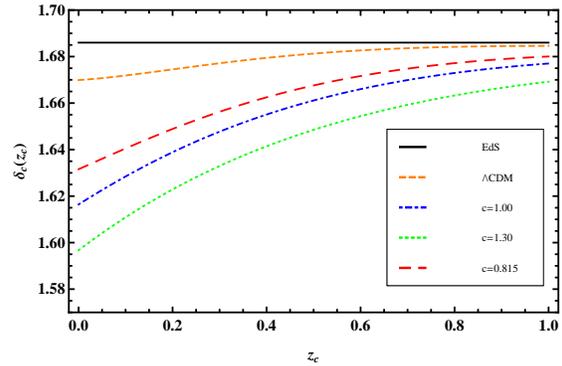}
 \caption{The variation of the linear threshold density contrast as a function of the collapse {\bf redshift} for the
 different cosmological background models analyzed in this work. Line styles and colors are as in Fig.~\ref{fig:gf}.}
 \label{fig:deltac}
\end{figure}

In this section, we present the numerical results of the spherical collapse model in HDE cosmologies and compare the
results with observations. In Fig.~(\ref{fig:deltac}) the evolution of the linear overdensity threshold parameter
$\delta_{\rm c}$ is shown as a function of redshift for different models. In the EdS model, $\delta_{\rm c}$ is
constant and redshift-independent as expected. In the $\Lambda$CDM model the linear overdensity parameter is
smaller than the corresponding value for the EdS model at low redshifts and it approaches the standard value of
$\delta_{\rm c}=1.686$ at higher redshifts. These two limits are easily explained. At high redshift the Universe is
matter dominated and therefore well described by an EdS model. At low redshift, the cosmological constant dominates,
therefore structures must form earlier with a lower critical density. In the case of the HDE model (phantom regime)
with constrained holographic parameter $c= 0.815$, $\delta_{\rm c}$ evolves more slowly than the $\Lambda$CDM model.
In the quintessence regime ($c\geq 1$) we see that the values of $\delta_{\rm c}$ are smaller than the others. These
behavior of $\delta_{\rm c}$ is explained by taking into account the evolution of the Hubble parameter in
Fig.~(\ref{fig:back}). The Hubble parameter is largest in the quintessence regime of HDE models, it takes
intermediate values in the phantom regime in HDE and smallest for $\Lambda$CDM model. Therefore the linear threshold
overdensity $\delta_{\rm c}$ for the HDE models will always have smaller values than in the $\Lambda$CDM universe.
Also note that the HDE cosmologies approach the EdS model at high redshifts, even if at a smaller rate than the
$\Lambda$CDM model.

\begin{figure}
 \centering
 \includegraphics[width=8cm]{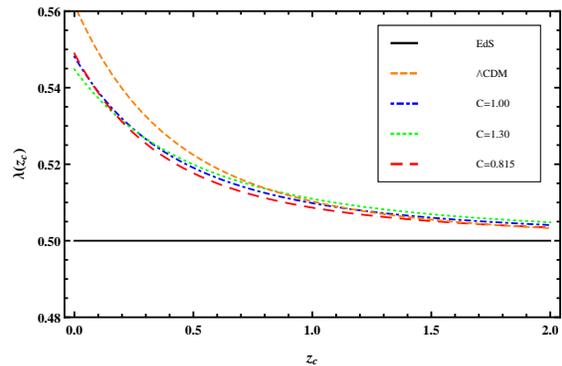}
 \caption{Dependency of the dimensionless parameter $\lambda=R_{\rm c}/R_{\rm ta}$ as a function of the collapse
 redshift $z_{\rm c}$ for different cosmological models. Line styles and colours are as in Fig.~\ref{fig:gf}.}
 \label{fig:lambda}
\end{figure}

Now by solving Eq.~(\ref{eqn:lambda}) we calculate the evolution of the dimensionless parameter
$\lambda=R_{\rm c}/R_{\rm ta}$ as a function of collapse redshift $z_{\rm c}$. In Fig.~(\ref{fig:lambda}),
we show $\lambda$ versus $z_{\rm c}$ for the HDE and $\Lambda$CDM models. We see that at redshifts large enough the
parameter $\lambda$ approaches the critical value $\lambda=1/2$ indicating once again that at large redshifts matter
dominates also in these models and the EdS model is a good approximation. This result is consistent with what found
for the linear overdensity parameter $\delta_{\rm c}$. Also note that, as for $\delta_{\rm c}$, the rate of
convergence to an EdS model is lower for HDE cosmologies than for a $\Lambda$CDM model. We can also conclude that the
size of structures in HDE models (for both phantom and quintessence regimes) are smaller than that predicted by the
$\Lambda$CDM model ($\lambda_{\rm HDE}\leq\lambda_{\Lambda\rm CDM}$).

\begin{figure}
 \centering
 \includegraphics[width=8cm]{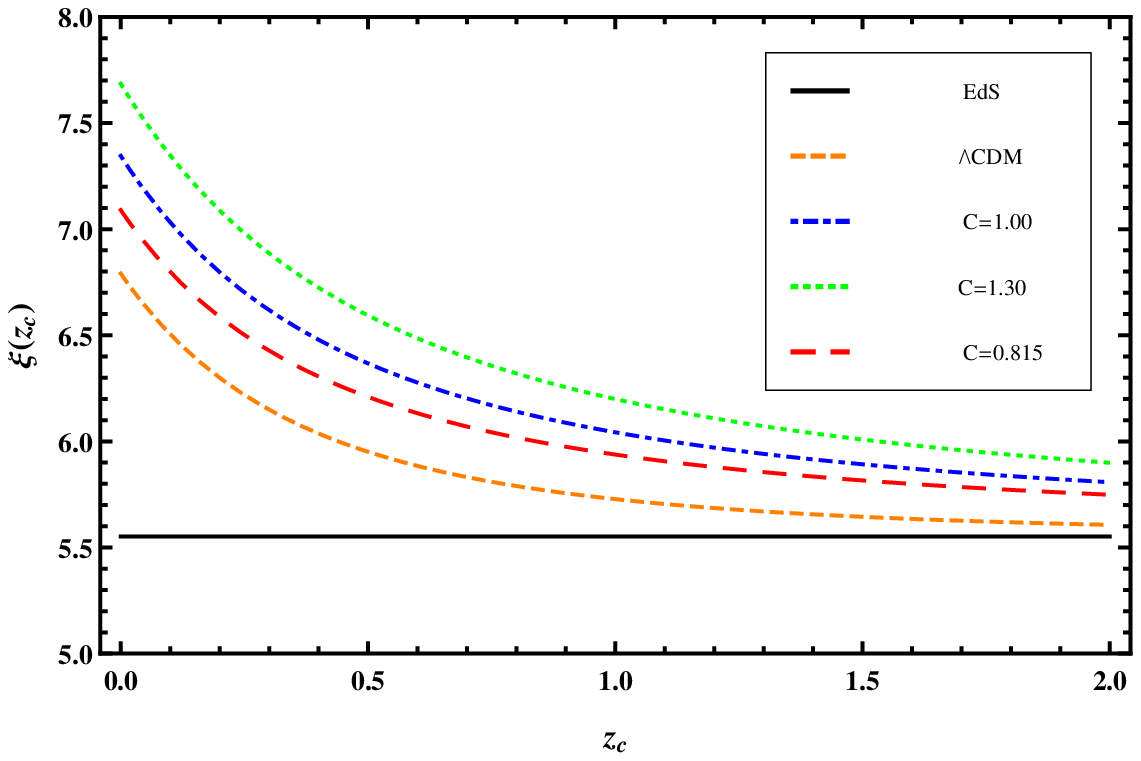}
 \includegraphics[width=8cm]{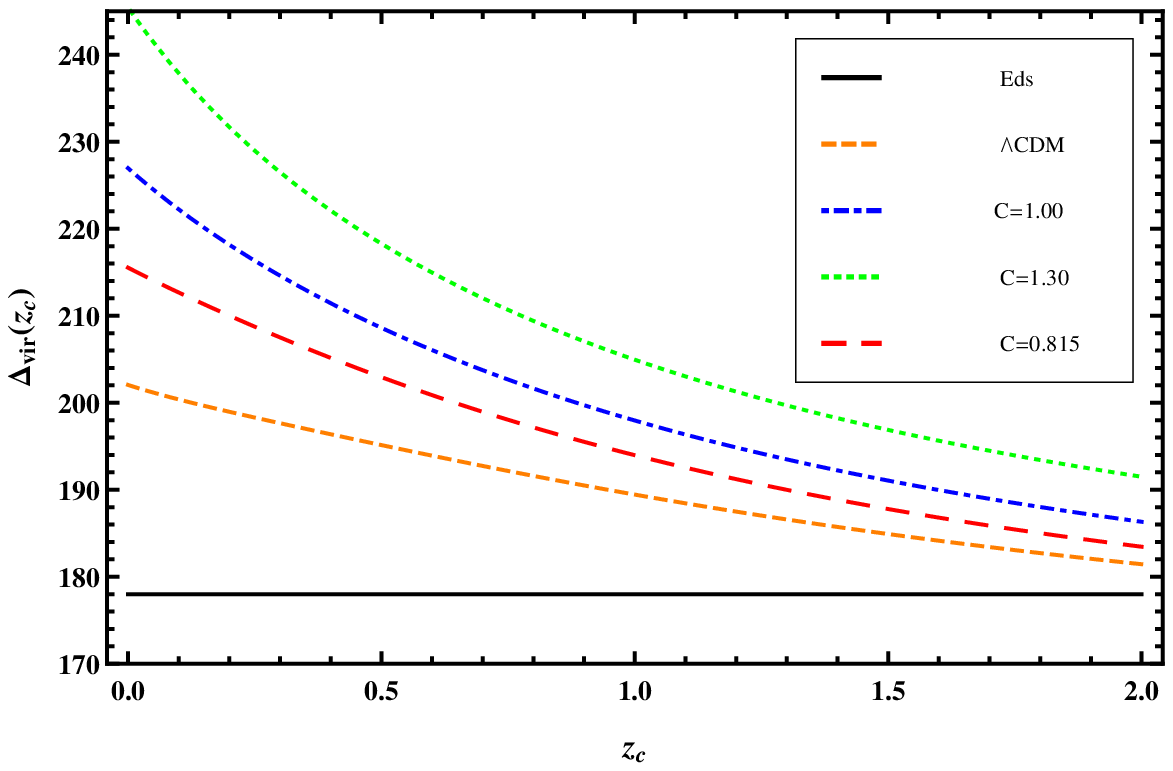}
 \caption{The variation of $\xi$ (top) and virial overdensity $\Delta_{\rm vir}$ (down) with collapse redshift
 $z_{\rm c}$ for various models considered in this work. Line styles and colours are as in Fig.~\ref{fig:gf}.}
 \label{fig:xi_DeltaV}
\end{figure}

In Fig.~(\ref{fig:xi_DeltaV}) the evolution of the parameter $\xi$ (top) and of the virial overdensity
$\Delta_{\rm vir}$ (down) with respect to the collapse redshift $z_{\rm c}$ is shown for different models.
At early times, $\xi$ converges to the fiducial value $\xi\approx 5.6$ which describes the early matter dominated
Universe. Compared to a $\Lambda$CDM model, the value of $\xi$ is larger for both quintessence and phantom regimes of
the HDE model. This means that the overdense spherical regions at turn-around are denser in HDE model than in
$\Lambda$CDM and EdS models. It is also interesting to notice that the overdensity at turn-around $\xi$ increases
monotonically as a function of the free parameter $c$. This is easily understood taking into account that with the
increase of $c$, the model switches from the phantom to the quintessence regime.

We finally discuss the evolution of the virial overdensity parameter $\Delta_{\rm vir}$. This quantity is important
for the definition of the halo size. Supposing that halos can be described approximately with a spherical geometry,
its mass $M_{\rm vir}$ and radius $R_{\rm vir}$ are linked by the following relation
\begin{equation}
 M_{\rm vir}=\frac{4\pi}{3}\rho_{\rm c}\Delta_{\rm vir}R_{\rm vir}^3\;,
\end{equation}
where $\rho_{\rm c}$ is the critical density of the Universe. In simulations studies, the critical density is often
replaced by the mean background density $\bar{\rho}_{\rm}=\rho_{\rm c}\Omega_{\rm m}$.

In analogy with what found for the overdensity at turn-around $\xi$, we see that the virial overdensity parameter
is higher in HDE cosmologies than for the $\Lambda$CDM model. Also coherently with previous results, this quantity
increases monotonically increasing the parameter $c$. As expected, differences are bigger at small redshifts
and gradually decrease at high redshift, where for sufficiently early times all the models will recover the EdS
solution, $\Delta_{\rm vir}\approx 178$. Phantom regime gives results closer to what predicted for the $\Lambda$CDM
cosmology. In this case ($c=0.815$) differences are of the order of 7\% and gradually increase up to more than 20\%
for $c=1.3$. Intermediate values are found for $c=1$.

\begin{figure}
 \centering
 \includegraphics[width=7.7cm]{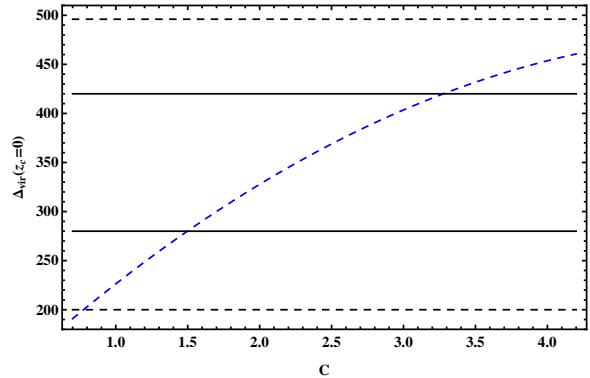}
 \caption{The present time virial overdensity for HDE model as a function of holographic parameter $c$. The
 horizontal solid and dashed lines correspond to the $1-$ and $2-\sigma$ observational overdensity limits
 respectively, based on a subsample of the 2MASS High Density Contrast group catalogue.}
 \label{fig:obs}
\end{figure}

We now compare $\Delta_{\rm vir}$ calculated in HDE cosmologies with the observational value of the virial
overdensity of clusters at the present time $\Delta_{\rm vir}^{\rm obs}(z_{\rm c}=0)$. To do this, we follow the work
of \cite{Basilakos2010} in which the present observational value for $\Delta_{\rm vir}^{\rm obs}(z_{\rm c}=0)$
with $1-\sigma$ ($2-\sigma$) error has been calculated as $348\pm 73(\pm 146)$ based on a sub sample of the 2MASS
high density contrast group catalogue \citep{Crook2007}. In Fig.~(\ref{fig:obs}) we plot the present time virial
density $\Delta_{\rm vir}$ in HDE models as a function of holographic parameter $c$. The horizontal solid and dashed
lines correspond to the $1-$ and $2-\sigma$ errors which limit the virial density to
$202\leq\Delta_{\rm vir}\leq 494$. This observational value for the present time sets limits on the value of the
holographic parameter to $0.8\leq c\leq4.5$.

\begin{figure}
 \centering
 \includegraphics[width=7.5cm]{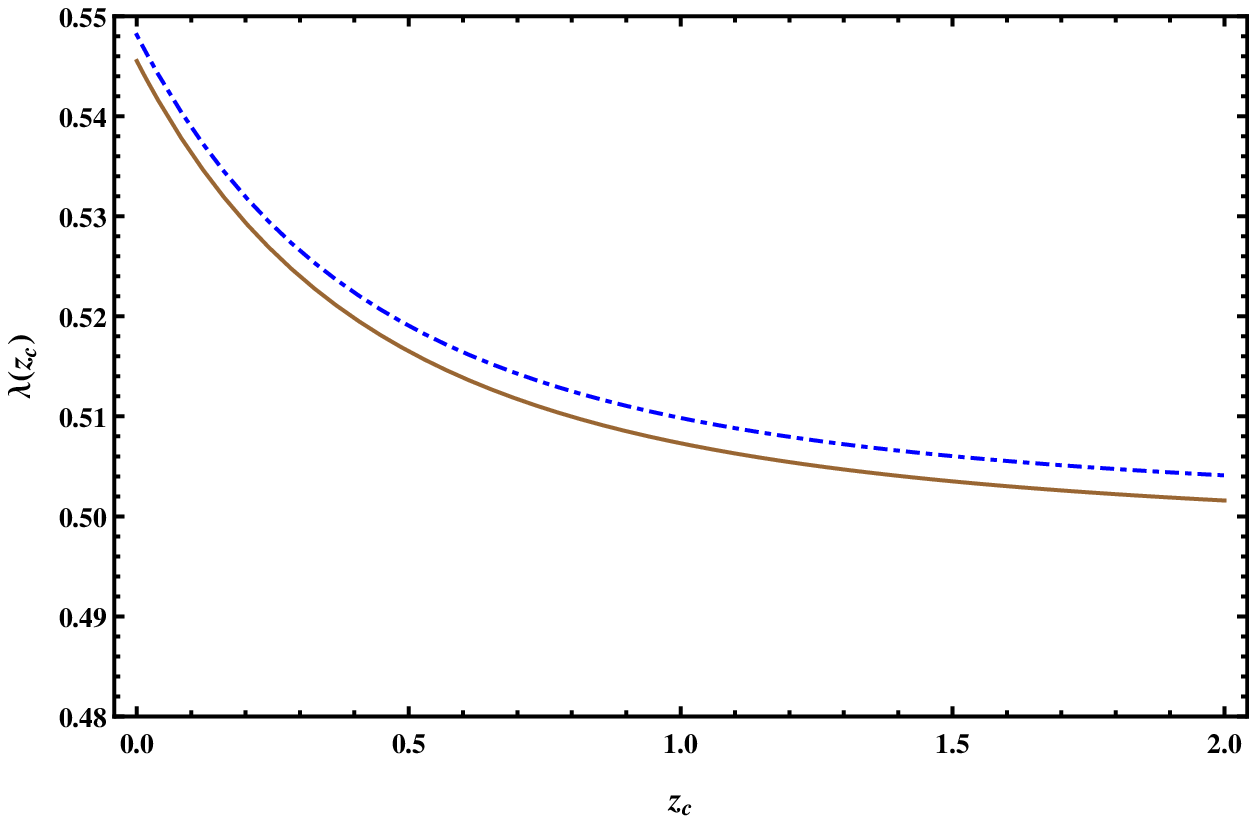}
 \includegraphics[width=7.5cm]{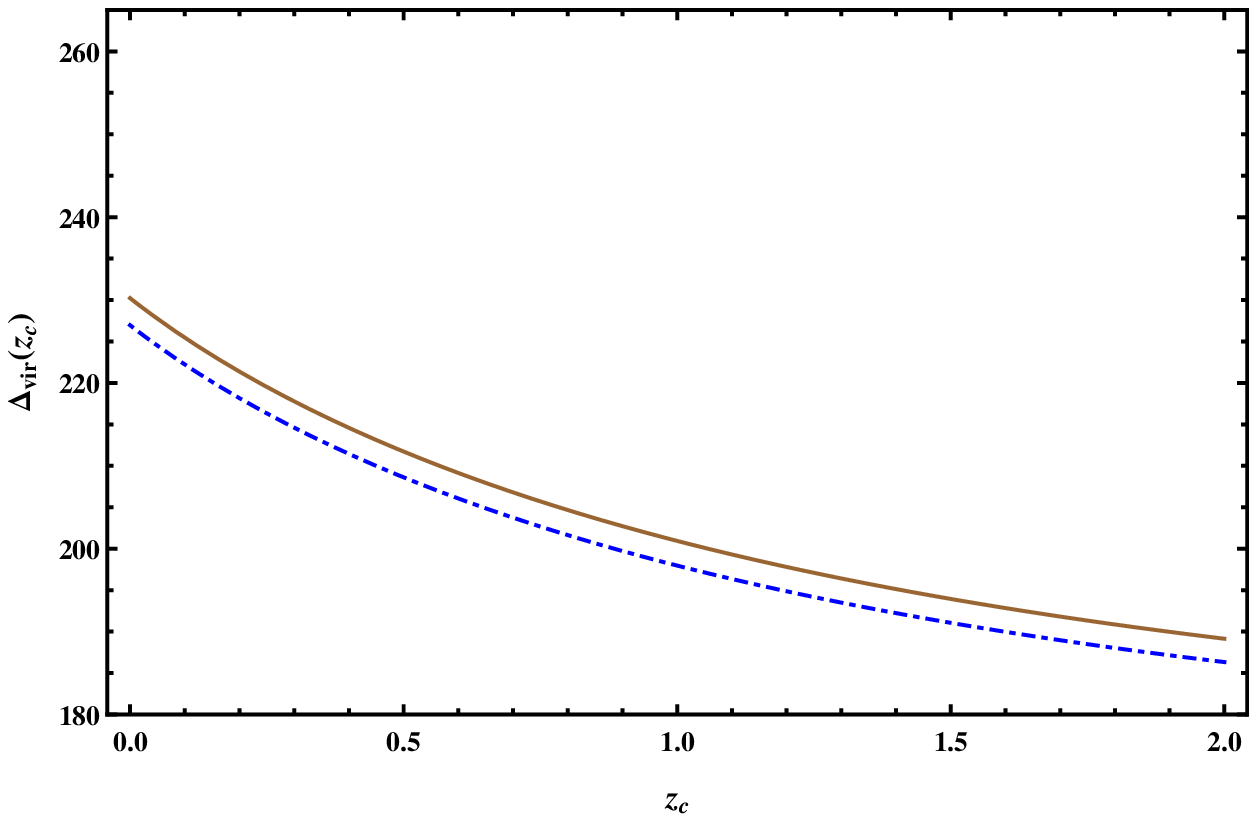}
 \caption{The variation of the dimensionless parameter $\lambda=R_{\rm c}/R_{\rm ta}$ (top) and virial overdensity
 $\Delta_{\rm vir}$ (down) as a function of collapse redshift $z_{\rm c}$ for HDE model ($c=1$) with non vanishing
 surface pressure (solid brown curve) and zero surface pressure (blue dashed curve).}
 \label{fig:pressure}
\end{figure}

 In Fig.~(\ref{fig:pressure}), we show the effect of the
non-vanishing surface pressure at the outer layers of clusters,
which appears due to non zero density at the boundary of clusters,
on the spherical collapse parameters $\lambda$ (top) and
$\Delta_{\rm vir}$ (down). The blue dashed (solid brown) curves
stand for HDE model with $c=1$ without (with) the effect of surface
pressure term. When including the surface pressure term we assume
$\nu=0.005$. We see that non vanishing surface pressure makes the
structures to virialize at a smaller radius with respect to standard
HDE models. Hence, the virial overdensity $\Delta_{\rm vir}$ is
higher by assuming surface pressure. Choosing the surface pressure
parameter $\nu=0.005$, differences between vanishing and
non-vanishing surface pressure HDE model ($c=1$) for parameter
$\lambda$ and $\Delta_{\rm vir}$ at the present time redshift
collapse ($z_c=0$) are of the order of $~1\%$ and $~1.5\%$,
respectively. Hence more denser and compact clusters can be formed
when including the effect of surface pressure.

 Using Eq.~(\ref{eq:sigma_8}), we obtain $\sigma_{8,\rm
HDE}(z=0)=0.76$ for $c=1.3$, $\sigma_{8,\rm HDE}(z=0)=0.77$ for
$c=1.0$ and $\sigma_{8,\rm HDE}(z=0)=0.78$ for $c=0.815$. Hence,
this quantity increases by decreasing the parameter $c$. For
$c=1.3$, differences from the $\Lambda$CDM value $\sigma_{8,\rm
\Lambda CDM}=0.8$ are of the order of $5\%$ and for $c=1$ and
$c=0.815$ are $4\%$ and $2.5\%$, respectively.

\begin{figure*}
 \begin{center}
  \includegraphics[width=8cm]{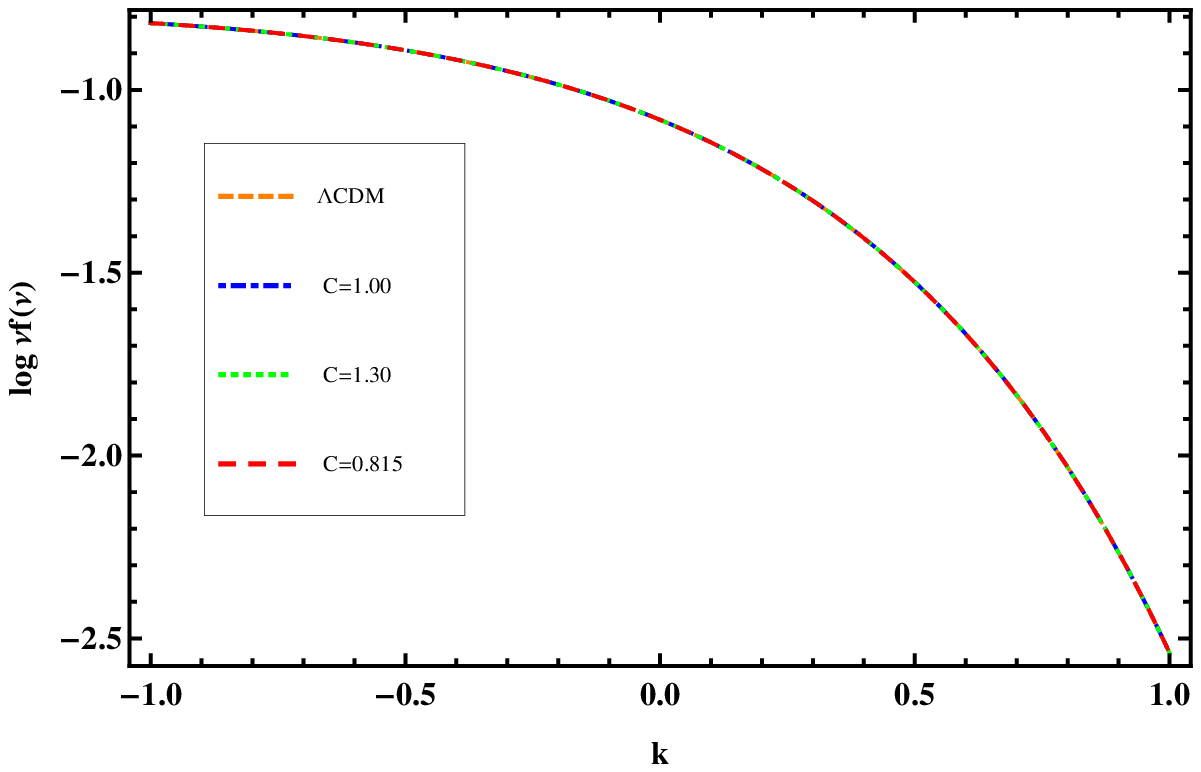}
  \includegraphics[width=8cm]{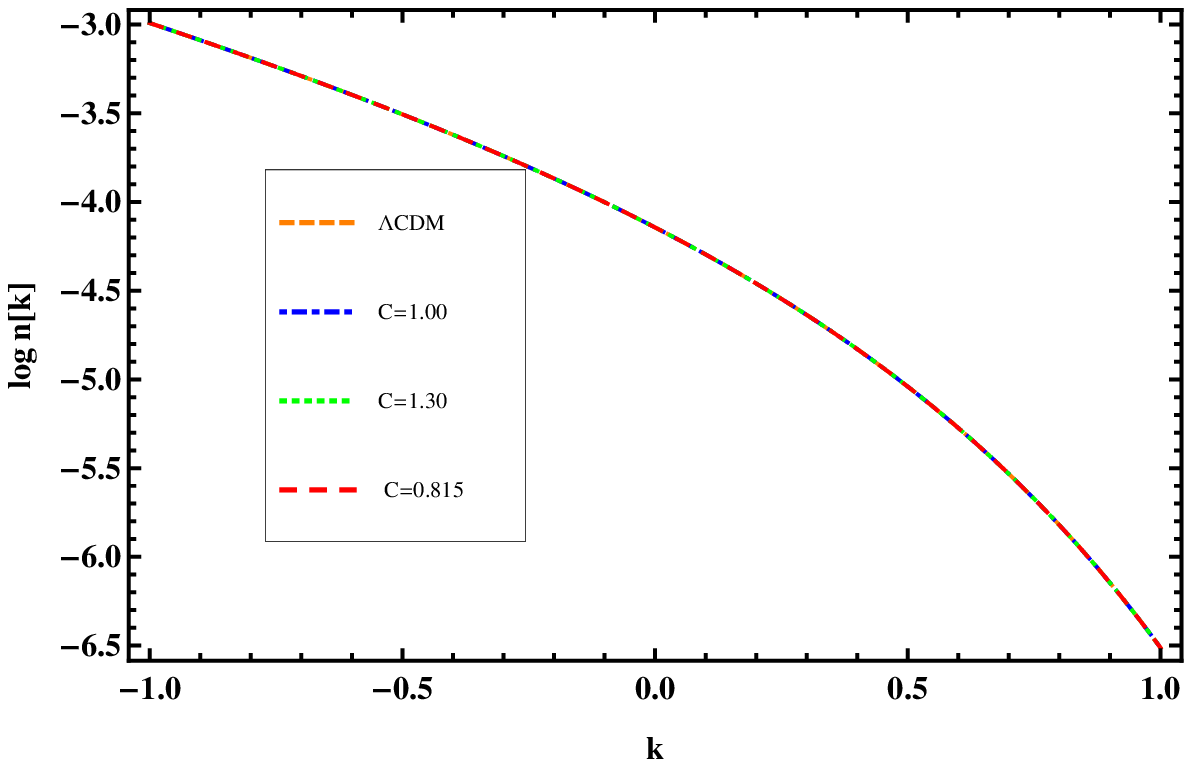}
  \includegraphics[width=8cm]{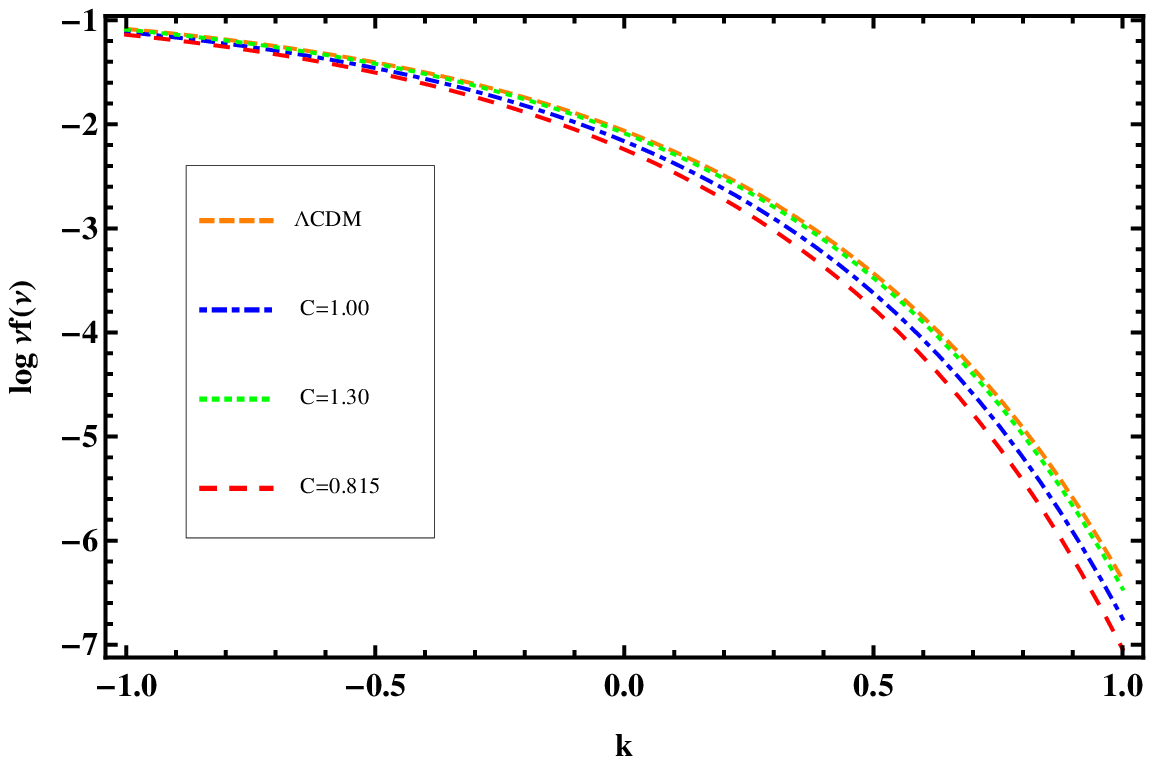}
  \includegraphics[width=8cm]{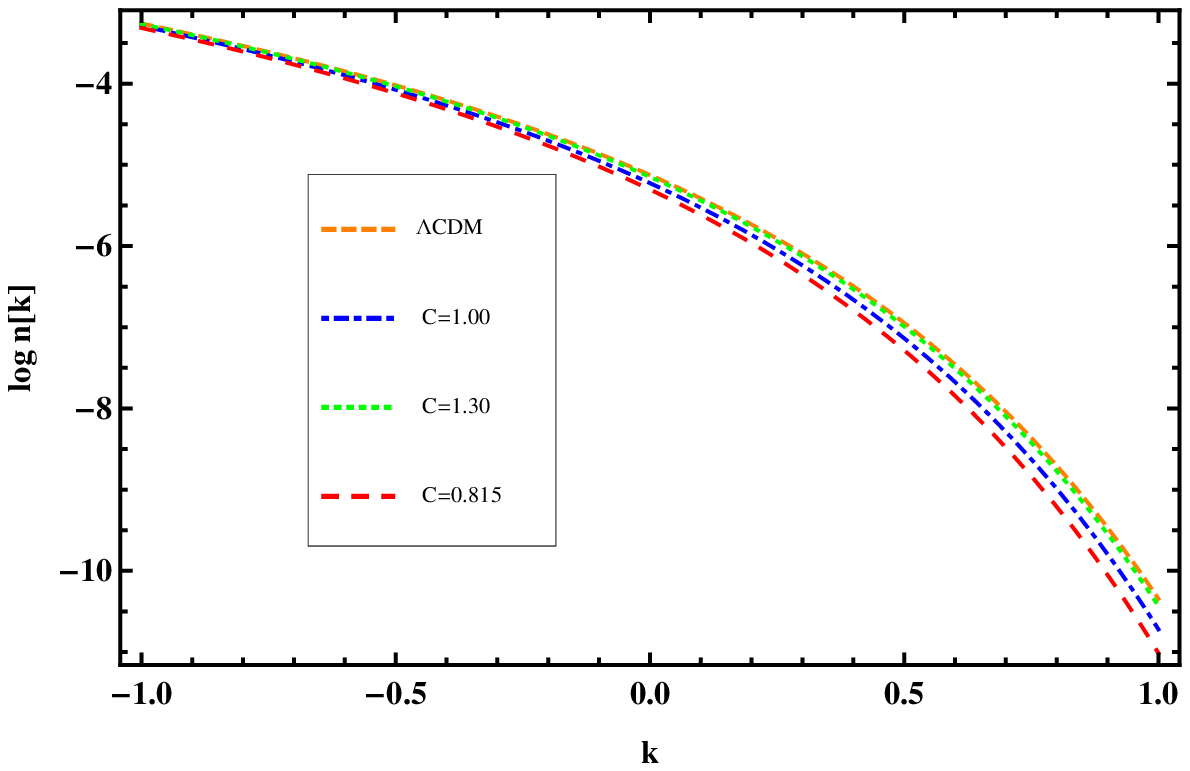}
  \caption{Mass function (left panels) and number density (right panels) for HDE and $\Lambda$CDM models at present
  time $z=0$ (upper panels) and at $z=1$ (lower panels), as described in the legend.}
  \label{fig:mass_fun}
 \end{center}
\end{figure*}

Fig.~(\ref{fig:mass_fun}) shows the evolution of the mass function $\nu f(\nu)$ in the upper left panel and the
number density $n(k)$, where $k=\log(M/M_8)$, in the upper right panel, as a function of $k$ for HDE cosmologies as
well as $\Lambda$CDM model at the present time by using the ST mass function in Eq.~(\ref{eq:multiplicity_ST})
proposed by Sheth \& Tormen \cite{Sheth1999,Sheth2002}. The lower panels show the same quantities, but at $z=1$. Due
to the normalization used, the mass function and number density (number of objects above a given mass) are the same
for all models at the present time $z=0$ and differences appear at high redshifts $z=1$. We also see that the
difference of the number densities of halo objects is in the high-mass tail and negligible for small objects
(see lower panel). We see in particular that decreasing the free parameter $c$, the number of objects per unit mass
and volume decreases. For values of $c>1$, the mass function for the HDE model is very similar to the $\Lambda$CDM
one, differing by it by only $1\%$. Since for $c<1$ we enter in the phantom regime, we expect this model to differ
mostly from the $\Lambda$CDM one. This is indeed the case as we can see in the lower panels. In this case differences
are of the order of $10\%$. Intermediate values take place for $c=1$ with differences of the order of $6\%$.

\begin{figure}
 \centering
 \includegraphics[width=7.7cm]{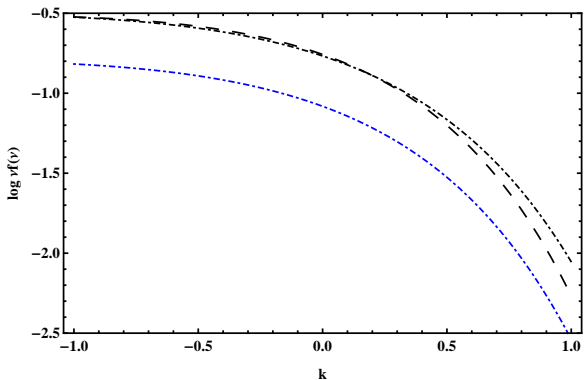}
 \caption{ Comparing various mass functions considered in this work for HDE model with $c=1$. The blue
 dotted-dashed curve indicates the ST mass function given by Eq.~(\ref{eq:multiplicity_ST}). The black dotted-dashed
 and dashed curves represent the improved mass functions P06 and YNY of Eqs.~(\ref{eqn:multiplicity_popolo2006} \&
 \ref{eqn:multiplicity_yahagi2004}), respectively.}
 \label{fig:mass_fun2}
\end{figure}

 Finally, we compare the results of ST mass function with the
improved versions P06 and YNY given in
Eqs.~(\ref{eqn:multiplicity_popolo2006} and
\ref{eqn:multiplicity_yahagi2004}), respectively, for HDE
cosmologies ($c=1$). We present our results in
Fig.~(\ref{fig:mass_fun2}). We see that the ST mass function (blue
dotted-dashed curve) is smaller than P06 (balck dotted-dashed) and
YNY (black dashed one), for all mass scales. We also see that P06
and YNY are the same for small mass scales and differ at high-mass
tail. Differences between the ST and the other mass functions are of
the order of 20\%, roughly constant throughout the whole mass range
investigated.

\section{Conclusions}\label{sect:conclusions}
In this work, we generalized the spherical collapse model in HDE
cosmologies with time varying EoS parameter $w_{\rm de}$. The
advantage of HDE models is that they are constructed on the basis of
the holographic principle in quantum gravity scenario
\citep{tHooft1993,Susskind1995}. We assumed two different phases of
HDE model i.e., quintessence regime ($c\geq1$) and phantom regime
($c<1$). In the case of phantom regime, we adopted the constrained
value for the model parameter $c=0.815$ obtained in
\cite{Enqvist2004,Gong2004,Huang2004a,Huang2004b,Li2009}. We first
investigated the growth of structures in linear regime and showed
that the growth of density perturbations $g(a)$ is slowed down in
HDE models compared to the EdS and $\Lambda$CDM models due to a
larger Hubble parameter. In particular, we showed that the growth
factor in quintessence and phantom regimes falls behind the
$\Lambda$CDM model (see Fig.~(\ref{fig:gf})). Therefore to observe
the same fluctuations at the present time, the perturbations should
start growing earlier in HDE model than in a $\Lambda$CDM model. We
then studied the non-linear phase of spherical collapse in HDE
model. We obtained fitting formulas governing the correlation
between turn-around redshift $z_{\rm ta}$ and virial redshift
$z_{\rm c}$ in HDE cosmologies. At large enough redshifts the effect
of DE in HDE models on turn-around is negligible and $z_{\rm ta}$
approaches the value of the EdS Universe. Using the generalized
virial condition obtained in time-varying HDE models, we obtained
the characteristic parameter of spherical collapse model
$\delta_{\rm c}$, $\xi$ (the overdensity at turn-around redshift)
and $\Delta_{\rm vir}$. We showed that the overdense spherical
region at the moment of turn-around are denser in HDE models than in
$\Lambda$CDM and EdS models. It has been shown that in HDE models
(both phantom and quintessence regimes), the virial overdensity
$\Delta_{\rm vir}$ is larger than in the $\Lambda$CDM model. Hence
in a HDE Universe more concentrated structures can be formed
compared to a $\Lambda$CDM model. Due to non-zero density at the
outer layers of clusters during the virialization process, we
studied the effect of a non-vanishing surface pressure on the
parameters of the spherical collapse in HDE cosmologies and showed
that in this case smaller virial radii and larger virial
overdensities $\Delta_{\rm vir}$ can be achieved with respect to the
case in which the surface pressure terms are neglected. We also
predicted that for larger values of the holographic parameter $c$
the virial overdensity $\Delta_{\rm vir}$ is larger. We could put a
limit on the holographic parameter $c$, based on the observational
value of present time virial overdensity $\Delta_{\rm vir}^{\rm
obs}(z_c=0)=348\pm 73(\pm 146)$ with $1-\sigma$ ($2-\sigma$)
error-bars calculated in \cite{Basilakos2010}. We showed that the
present time virial overdensity in HDE cosmology has a good
agreement with observational value at $2-\sigma$ level for the
holographic parameter $c$ in the range of $0.8\leq c\leq4.5$.
Finally, using the fitting formula proposed by Sheth \& Tormen
\citep{Sheth1999,Sheth2002} for the mass function, we obtained the
mass function and number densities of clusters in HDE cosmologies.
The number densities of small halo objects are the same for all
models. However, we showed that the HDE models deviate from
concordance $\Lambda$CDM model at high mass halo objects. We showed
that the ST mass functions in HDE cosmologies is smaller than the
improved mass functions presented in \cite{Popolo2006a,Popolo2006b}
and \cite{Yahagi2004} for all mass scales investigated.

\section*{Acknowledgements}
We would like to thank an anonymous referee for giving us constructive comments that helped us to improve the
scientific content of the manuscript.\\
Francesco Pace is supported by STFC grant ST/H002774/1.\\
T. Naderi and M. Malekjani thank A. Mehrabi for useful discussions.

\bibliographystyle{mn2e}
\bibliography{ref}

\label{lastpage}

\end{document}